\documentclass[floatfix,aps,prb,10pt,showpacs,twocolumn,superscriptaddress]{revtex4}
\usepackage{xcolor}
\usepackage{amsmath}
\usepackage{amssymb}
\usepackage{graphicx}
\usepackage[colorlinks,linkcolor=blue,urlcolor=blue,citecolor=blue]{hyperref}
\usepackage[all]{hypcap} % let hyperlinks correctly point to figures rather than their captions; must be loaded after the hyperref package!
\begin{document}

%%%% Some macros
\newcommand\hatc{\hat{c}}
\newcommand\hatf{\hat{f}}
\newcommand\hatH{\hat{H}}
\newcommand\calG{\mathcal{G}}
\newcommand\calH{\mathcal{H}}
\newcommand\calS{\mathcal{S}}
\newcommand\msc[1]{\textcolor{red}{\textbf{MSC:} #1}}

\title{Subgap modes in two-dimensional magnetic Josephson junctions}

\author{Yinan Fang}
\affiliation{School of Physics and Astronomy, Yunnan University, Kunming 650091, China}

\author{Seungju Han}
\affiliation{Department of Physics, Korea University, Seoul 02841, South Korea}

\author{Stefano Chesi}
\email{stefano.chesi@csrc.ac.cn}
\affiliation{Beijing Computational Science Research Center, Beijing 100193, People's Republic of China}
\affiliation{Department of Physics, Beijing Normal University, Beijing 100875, People's Republic of China}

\author{Mahn-Soo Choi}
\email{choims@korea.ac.kr}
\affiliation{Department of Physics, Korea University, Seoul 02841, South Korea}

\begin{abstract}
We consider two-dimensional superconductor/ferromagnet/superconductor junctions and investigate the subgap modes along the junction interface.
The subgap modes exhibit characteristics similar to the
Yu-Shiba-Rusinov states that originate form the interplay between superconductivity and ferromagnetism in the magnetic junction.
The dispersion relation of the subgap modes shows qualitatively different profiles depending on the transport state (metallic, half-metallic, or insulating) of the ferromagnet.
As the spin splitting in the ferromagnet
is increased, the subgap modes bring about a $0$-$\pi$ transition in the Josephson current across the junction, with the Josephson current density depending strongly on the momentum along the junction interface (i.e., the direction of the incident current).
For clean superconductor-ferromagnet interfaces (i.e., strong coupling between superconductors and ferromagnet), the subgap modes develop flat quasi-particle bands that allow to engineer the wave functions of the subgap modes along an inhomogeneous magnetic junction.
\end{abstract}


\maketitle

%%%% To be removed later
% \tableofcontents

\section{Introduction}
\label{SFS2D::sec:1}

The interplay between superconducting and ferromagnetic order leads to
unconventional pairing mechanisms as well as exotic quantum states, such as
the Yu-Shiba-Rusinov (YSR) state bounded to a (classical) magnetic impurity,\cite{1965_Acta_Phys_Sin_Yu, 1968_Prog_Theor_Phys_Shiba, 1969_JETP_Lett_Rusinov}
the Fulde-Ferrell-Larkin-Ovchinnikov (FFLO) states in ferromagnetic metals,\cite{1964_PR_Flude,1965_JETP_Larkin} and
the chiral Majorana edge modes in topological superconductors.\cite{He17a,Qi10b}
Understanding and controlling the delicate competition between different
orders will undoubtedly benefit the development of quantum devices for various
spintronic applications.\cite{2011_Phys_Today_Eschrig}

In a Josephson junction, the transport properties are governed by subgap
states below the superconducting energy gap, and these subgap states
reflect the fate of the competition between superconductivity and magnetism.
For example, consider a Josephson junction through a quantum dot,\cite{Kouwenhoven01a} which can be regarded as a magnetic impurity with strong quantum fluctuations.
The subgap state induced by the impurity behaves like an Andreev bound state in the strong-coupling limit, where the Kondo effect\cite{1964_ProgTheorPhys_Kondo, 1975_RMP_Wilson, 1993_BOOK_Hewson} dominates over superconductivity,
whereas it bears a closer resemblance to the YSR state in the weak-coupling limit,\cite{1970_ProgTheorPhys_Sakurai, 2006_RMP_Balatsky} where superconductivity dominates over the Kondo effect.
Such change in character of the subgap state results in 
a transition from negative to positive supercurrent across the junction, usually referred to as a quantum phase transition from  a $0$-junction to a $\pi$-junction.\cite{2004_PRB_Choi,2010_PRL_Zitko}

In a superconductor/ferromagnet/superconductor (S/FM/S) junction, the nature
of subgap states depends on the transport properties of the ferromagnetic
layer.
When the ferromagnetic layer is metallic, spin-dependent Andreev subgap states play a dominant role: The finite center-of-mass
momentum of Cooper pairs which penetrate into the ferromagnetic metal
causes an oscillatory behavior in the proximity-induced pairing potential.\cite{1991_PRB_Andreev,2011_Phys_Today_Eschrig}
Depending on the relative width of the ferromagnetic layer with respect to the
wave length of the oscillation, the ground state of the S/FM/S junction may be
stabilized with either a $0$ or $\pi$ phase difference between the two
superconductors.\cite{1977_JETPLett_Bulaevskii, 1982_JETPLett_Buzdin, 2001_PRL_Ryazanov_Coupling}
In a recent paper,\cite{2018_PRB_Costa} however, it was found that the YSR subgap states play a more
significant role when the ferromagnet is a thin insulator. The competition of superconductivity versus
magnetism induces a strong dependence of the YSR state on the spin splitting
in the ferromagnet, leading to a $0$-$\pi$ transition in the junction when the spin splitting is increased.\cite{2018_PRB_Costa}

\begin{figure}
\begin{centering}
\includegraphics[width=8cm]{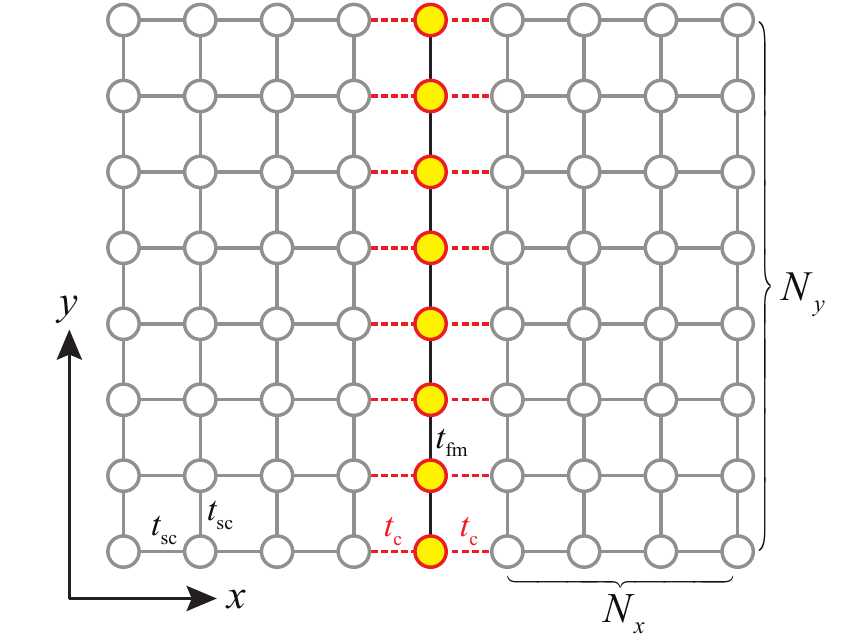}
\par\end{centering}
\caption{(color online) A schematic tight-binding model for the
  two-dimensional superconductor/ferromagnet/superconductor junction.  The
  empty gray circles represent the lattice sites on the left and right
  superconductors, whereas the red circles filled in yellow denote sites in
  the ferromagnet.  The homogeneous nearest-neighbor tunnel coupling strengths
  within the superconducting and ferromagnetic regions ($t_\mathrm{sc}$ and
  $t_\mathrm{fm}$) are denoted by gray and black solid links, respectively,
  and the coupling between superconductors and ferromagnet ($t_\mathrm{c}$) is
  depicted by red dashed links.  The system consists of $2N_{x}+1$ sites along
  the $x$-direction and $N_{y}$ sites along the
  $y$-direction.\label{FIG_junction_schematics}}
\end{figure}

While so far most of the previous works studied one-dimensional (1D) or quasi-1D junctions, i.e., narrow junctions,
in this work we consider two-dimensional S/FM/S junctions (Fig.~\ref{FIG_junction_schematics})
and investigate the subgap modes along the junction interface.
We find that, due to the interplay between superconductivity and ferromagnetism in the magnetic junction, the subgap modes inherit the characteristics of the 
YSR states  and lead to the following intriguing properties that are hard to observe in narrow junctions:
(i) The dispersion relation of the subgap modes shows qualitatively different profiles depending on the transport state (metallic, half-metallic, or insulating) of the ferromagnet.
(ii) The subgap modes mediate the $0$-$\pi$ transition in the Josephson
current across the junction, induced by increasing the spin splitting in the ferromagnet. They also determine a dependence of the Josephson current density on the superconducting
phase difference which changes sharply with the momentum along the junction
interface (i.e., the direction of the incident current).
(iii) For clean superconductor-ferromagnet interfaces (i.e., strong coupling between superconductors and ferromagnet), the subgap modes develop flat quasi-particle bands that allow to engineer the wave functions of the subgap modes along an inhomogeneous magnetic junction.

Apart from these results, which are interesting from a fundamental point of view,
we note that several recent studies used scanning tunneling
microscopy/spectroscopy to explore the exotic physics associated with
chains of magnetic atoms.\cite{2014_Science_NadjPerge, 2015_PRL_Peng,
  2015_NatCommun_Hatter, 2021_PNAS_Ding} This suggests that the characterization of S/FM/S junction of intermediate size,
interpolating between the few impurities and the continuum junction limits, can be useful also for device
applications.

The outline of the rest of our paper is as follows: In Section~\ref{SFS2D::sec:2}, we present
the model of the S/FM/S junction. In Section~\ref{SFS2D::sec:3}, we discuss the characteristic properties of the subgap modes as, in particular, their dispersion relation and dependence on the superconducting phase difference.
Section~\ref{SFS2D::sec:5} is devoted to the $0$-$\pi$ transition of the magnetic Josephson junction and Section~\ref{SFS2D::sec:4} discusses the quasi-particle flat bands of the subgap modes, together with their effect on the wave functions of the subgap modes along inhomogeneous magnetic junctions.
Section~\ref{SFS2D::sec:6} concludes the paper.

\section{Models}
\label{SFS2D::sec:2}

We consider a two-dimensional (2D) S/FM/S junction, shown schematically in Fig.~\ref{FIG_junction_schematics}, and
describe it with the following tight-binding Hamiltonian\cite{1996_JPSJ_Onishi,2000_PRB_Zhu}
\begin{equation}
\label{Hamiltonian}
\hat{H}=\hat{H}_{\mathrm{L}}+\hat{H}_{\mathrm{R}}+\hat{H}_{\mathrm{fm}}+\hat{H}_{\mathrm{tun}} .
\end{equation}
Terms $\hat{H}_{\mathrm{L/R}}$ are responsible for the left (L) and
right (R) superconductors, respectively, and are given by
\begin{align}
\label{SC_Hamiltonian}
\hat{H}_{\mathrm{L/R}}= &
-\frac{t_\mathrm{sc}}{2}
\sum_{x=1}^{N_{x}}\sum_{y=1}^{N_{y}}\sum_{\sigma=\uparrow,\downarrow}
\hatc_{\mp x,y,\sigma}^\dag
\hatc_{\mp x,y+1,\sigma} + \mathrm{h.c.}
\nonumber\\ & 
-\frac{t_\mathrm{sc}}{2}
\sum_{x=1}^{N_{x}-1}\sum_{y}\sum_{\sigma}
\hatc_{\mp x,y,\sigma}^\dag
\hatc_{\mp x+1,y,\sigma} + \mathrm{h.c.}
\nonumber\\ & 
+ \Delta_{\mathrm{L/R}}\sum_{x=1}^{N_x}\sum_y
\hat{c}_{\mp x,y,\uparrow}^{\dagger}
\hat{c}_{\mp x,y,\downarrow}^{\dagger} + \mathrm{h.c.}
\nonumber \\ & 
- \mu_\mathrm{sc}\sum_{x=1}^{N_x}\sum_y\sum_\sigma
\hatc_{\mp x,y,\sigma}^\dag\hatc_{\mp x,y,\sigma} \,,
\end{align}
where $\hatc_{x,y,\sigma}$ is the electron annihilation operator for spin $\sigma$ at site $(x,y)$ of the superconductors,
$\mu_\mathrm{sc}$ is the Fermi energy of the superconductors, and
\begin{math}
\Delta_{\mathrm{L}/\mathrm{R}}
= \Delta_{\mathrm{sc}} \exp[i\varphi_{\mathrm{L}/\mathrm{R}}]
\end{math}
($\Delta_\mathrm{sc}>0$)
is the superconducting pairing potential of each superconductor.
Here, we impose periodic boundary conditions in the $y$-direction (%
\begin{math}
\hatc_{x,N_y+1,\sigma} = \hatc_{x,1,\sigma}
\end{math}
for all $x$ and $\sigma$), and open boundary conditions in the $x$-direction. We also
assume identical superconductors on both sides of the junction
\begin{math}
(|\Delta_\mathrm{L}|=|\Delta_\mathrm{R}|).
\end{math}
In the following, we measure energy from the Fermi level and denote with $\varphi = \varphi_\mathrm{L} - \varphi_\mathrm{R}$ the phase across the junction. Term $\hat{H}_{\mathrm{fm}}$ describes the one-dimensional ferromagnet along the $y$-direction (see Fig. \ref{FIG_junction_schematics}) and is given by
\begin{multline}
\label{SFS2D::eq:3}
\hat{H}_{\mathrm{fm}}= \sum_{y}\left[\lambda_{\mathrm{M}}\left(\hat{f}_{y,\uparrow}^{\dagger}\hat{f}_{y,\uparrow}-\hat{f}_{y,\downarrow}^{\dagger}\hat{f}_{y,\downarrow}\right)-\mu_{\mathrm{fm}}\sum_{\sigma}\hat{f}_{y,\sigma}^{\dagger}\hat{f}_{y,\sigma}\right.
\\ \left.-\frac{t_{\mathrm{fm}}}{2}\sum_{\sigma}\left(\hat{f}_{y,\sigma}^{\dagger}\hat{f}_{y+1,\sigma}+\mathrm{h.c.}\right)\right],
\end{multline}
where $\hatf_{y,\sigma}$ is the electron annihilation operator for spin $\sigma$ at site $y$  (recall the periodic boundary condition, $\hatf_{N_y+1,\sigma}=\hatf_{1,\sigma}$), $\mu_\mathrm{fm}$ is the Fermi energy of the ferromagnet, 
and $\lambda_{\mathrm{M}}$ is the magnetic spin-splitting due
to the ferromagnetic order.\footnote{Neglecting the dynamic effect associated with the magnetic moment
is valid if the magnetic moment is sufficiently large and the temperature is well below the Kondo temperature\cite{2006_RMP_Balatsky}.}
Finally, the last term $\hat{H}_{\mathrm{tun}}$ describes the tunnel
coupling between the ferromagnet and superconductors, that is,
\begin{equation}
\hat{H}_{\mathrm{tun}}=-\frac{t_\mathrm{c}}{2}\sum_{x=\pm1}\sum_{y}\sum_{\sigma}\left(\hat{c}_{x,y,\sigma}^{\dagger}\hat{f}_{y,\sigma}+\mathrm{h.c.}\right).
\end{equation}
% In above, $\hat{c}_{x,y,\sigma}$ and $\hat{f}_{y,\sigma}$ are the electron
% annihilation operators of spin $\sigma$ in SCs and ferromagnetic regions,
% respectively.

%%%%%%%%%%%%%%%%%%%%%%%%%%%%%%%%%%%%%%%%%%%%%%%%%%%%
\begin{figure*}
\begin{centering}
\includegraphics[width=15cm]{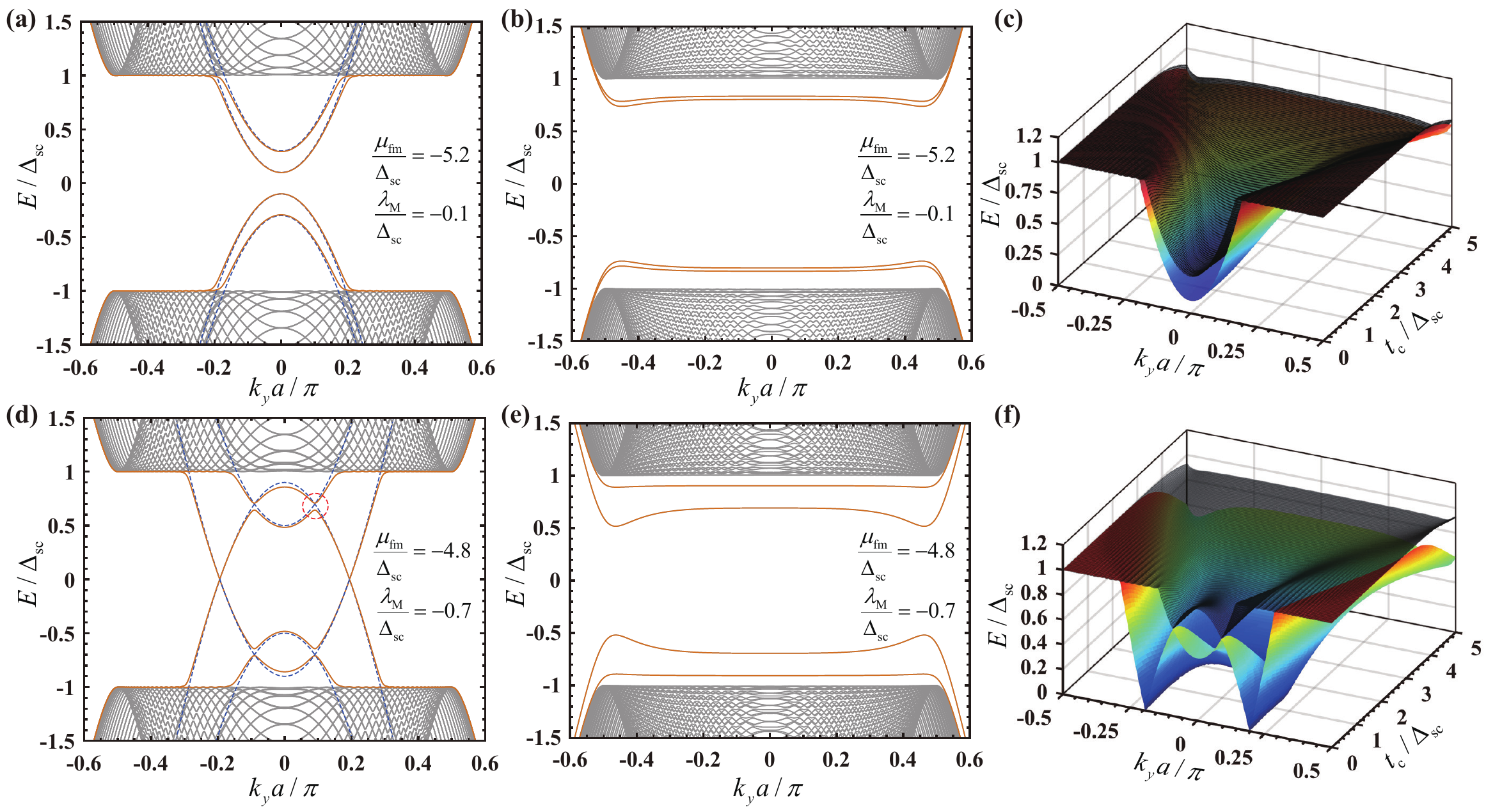}
\par\end{centering}
\caption{(color online) Energy spectrum of the subgap modes in the S/FM/S
  Josephson junction as a function of the transverse wave number $k_{y}$ and
  tunnel coupling $t_c$ between the superconductor and ferromagnet.  The upper
  panels (a--c) are the S/FM/S Josephson junction with an insulating
  ferromagnet and the lower panels (d--f) are for the junction with a
  half-metallic ferromagnet.  In (a) and (d), the dispersions of isolated
  ferromagnets are shown as dashed curves for reference.  The anti-crossing
  marked by the red circle in (d) is further analyzed in
  Fig.~\ref{FIG_EffectivePairingPotential}; see also the main text. We used
  $\mu_\mathrm{fm}=-5.2\Delta_\mathrm{sc}$ and
  $\lambda_\mathrm{M}=-0.1\Delta_\mathrm{sc}$ in the upper panels and
  $\mu_\mathrm{fm}=-4.8\Delta_\mathrm{sc}$ and
  $\lambda_\mathrm{M}=-0.7\Delta_\mathrm{sc}$ in the lower panels.  Other
  parameters: $N_{x}=60$, $\varphi=\pi/4$,
  $\mu_{\mathrm{sc}}=-5\Delta_{\mathrm{sc}}$, and
  $t_{\mathrm{sc}}=t_{\mathrm{fm}}=5\Delta_{\mathrm{sc}}$. The
  superconductor-ferromagnetic coupling is
  $t_\mathrm{c}=0.4\Delta_{\mathrm{sc}}$ for panels (a,d) and
  $t_\mathrm{c}=5\Delta_{\mathrm{sc}}$ for panels (b,e).}
\label{FIG_ThreeRegimes}
\end{figure*}
%%%%%%%%%%%%%%%%%%%%%%%%%%%%%%%%%%%%%%%%%%%%%%%%%%%%

Throughout the paper, we will mainly discuss the numerical results based on
the above tight-binding model. On the other hand, it turns out that some of
the results may be understood more transparently with a continuum model
\cite{1982_PRB_Blonder,1991_PRL_Beenakker} governed by the Bogoliubov de
Gennes (BdG) Hamiltonian \cite{1966_BOOK_deGennes,2014_PRB_Zyuzin,2021_PRB_Deb}
\begin{equation}
\label{SFS2D::eq:1}
\hat{H}=\frac{1}{2}\int d^{2}\mathbf{r}\,
\hat{\Psi}(\mathbf{r})^{\dagger}
\mathcal{H}_{\mathrm{BdG}}(\mathbf{r})\hat{\Psi}(\mathbf{r})
\end{equation}
for a magnetic Josephson junction between two superconductors ($|x|>L/2$) through a narrow ferromagnet ($|x|<L/2$)
with
\begin{align}
\mathcal{H}_{\mathrm{BdG}}(\mathbf{r})= &
\left[-\frac{\hbar^{2}\nabla^{2}}{2m}-\mu(x)+V_{b}(x)\right]
\tau^{z}\sigma^{0}
\nonumber \\ & 
+ \Delta_{\mathrm{sc}}
\theta(x-L/2)\left(\cos\varphi_{\mathrm{R}}\tau^{x}-\sin\varphi_{\mathrm{R}}\tau^{y}\right)\sigma^{0}
\nonumber \\ & 
+ \Delta_{\mathrm{sc}}
\theta(-x-L/2)\left(\cos\varphi_{\mathrm{L}}\tau^{x}-\sin\varphi_{\mathrm{L}}\tau^{y}\right)\sigma^{0}
\nonumber \\ & 
+\lambda_\mathrm{M}\theta(L/2-|x|)\tau^{0}\sigma^{z} ,
\label{BdG_Hamiltonian}
\end{align}
where
$\hat{\Psi}(\mathbf{r})$ is the Nambu spinor,
\begin{equation}
\label{Nambu_convention}
\hat{\Psi}(\mathbf{r})=\left[\begin{array}{cccc}
\hat{\psi}_{\uparrow}(\mathbf{r}) & \hat{\psi}_{\downarrow}(\mathbf{r}) &
\hat{\psi}_{\downarrow}(\mathbf{r})^{\dagger} &
-\hat{\psi}_{\uparrow}(\mathbf{r})^{\dagger}\end{array}\right]^{\mathrm{T}},
\end{equation}
and $\sigma^{\alpha}$ ($\tau^{\alpha}$) for $\alpha=0,1,2,3$ represent the Pauli matrices acting on the spin (particle-hole) subspace ($\alpha=0$ corresponds to the identity matrix).
In Eq.~(\ref{BdG_Hamiltonian}), the chemical potential takes two different values for the superconductors ($\bar\mu_\mathrm{sc}$) and the ferromagnet ($\bar\mu_\mathrm{fm}$), i.e., 
\begin{equation}\label{mu_continuum_model}
\mu(x) =
\begin{cases}
\bar{\mu}_\mathrm{sc}, & |x|>L/2\,, \\
\bar{\mu}_\mathrm{fm}, & |x|<L/2\,,
\end{cases}
\end{equation}
and  $\lambda_\mathrm{M}$ quantifies the magnetic spin splitting of the ferromagnet.
The interface between the ferromagnet and superconductors is modeled by $\delta$-function tunnel barriers
\begin{equation}\label{Vb}
V_{b}(x)=l_0 V_{0}\delta(x-L/2)+l_0 V_{0}\delta(x+L/2),
\end{equation}
where $l_0$ is an arbitrary parameter, physically characterizing the width of the (thin) potential barrier. In Appendix~\ref{SFS2D::sec:B}, we discuss the correspondence between continuum and tight-binding models and obtain that, when $t_\mathrm{c}\ll t_\mathrm{sc}\approx t_\mathrm{fm}$, the strength $l_0 V_0$ of the interface potential can be related to the tunneling amplitudes $t_\mathrm{sc}, t_\mathrm{c}$ of Eq.~\eqref{Hamiltonian} as follows:
\begin{align}
\frac{1}{\pi}
\left(\frac{t_{\mathrm{sc}}}{t_\mathrm{c}}
-\frac{t_\mathrm{c}}{t_{\mathrm{sc}}}\right) &
\approx \sqrt{\frac{2m}{\bar\mu_\mathrm{sc}}}\frac{\lambda_0 V_{0}}{\hbar}.
\label{SFS2D::eq:2}
\end{align}
Here, we have assumed that the Fermi wavenumber and the lattice constant $a$ of the tight-binding model satisfy
\begin{math}
k_\mathrm{F} a\approx\pi,
\end{math}
where $k_\mathrm{F}=\sqrt{2m\bar{\mu}_\mathrm{sc}}/\hbar$.
In Appendix~\ref{SFS2D::sec:B}, we also consider the relation between chemical potentials entering the two models. In particular, the difference between $\mu_\mathrm{fm}$ and $\bar{\mu}_\mathrm{fm}$ depends on the confinement energy in the ferromagnetic strip, that is affected in a nontrivial way by both $L$ and the strength of interface terms.

\section{Subgap Modes}
\label{SFS2D::sec:3}

When a quasi-particle has energy lower than the superconducting gap, it cannot penetrate into the superconductors and only propagates inside the junction, that is, within the ferromagnetic region of the S/FM/S junction.
Nonetheless, in a wide junction, the quasi-particle can move along the interface direction
(i.e., the $y$-direction in Fig. \ref{FIG_junction_schematics}) which, for the moment, we treat as translationally invariant.
Under this assumption, subgap states of quasi-particles form one-dimensional traveling modes with wave number $k_y$.
In this section, we investigate the characteristic properties of the subgap modes.

While subgap states in S/FM/S junctions have already been studied previously,\cite{2018_PRB_Costa,2019_PRB_Rouco}
they were focused on the one-dimensional (1D) or quasi-1D limit. For the wide junctions of our concern, the additional dimension along the junction interface should be explicitly accounted for, allowing for richer phenomena in the subgap region.

\begin{figure}
\begin{centering}
\includegraphics[width=8.6cm]{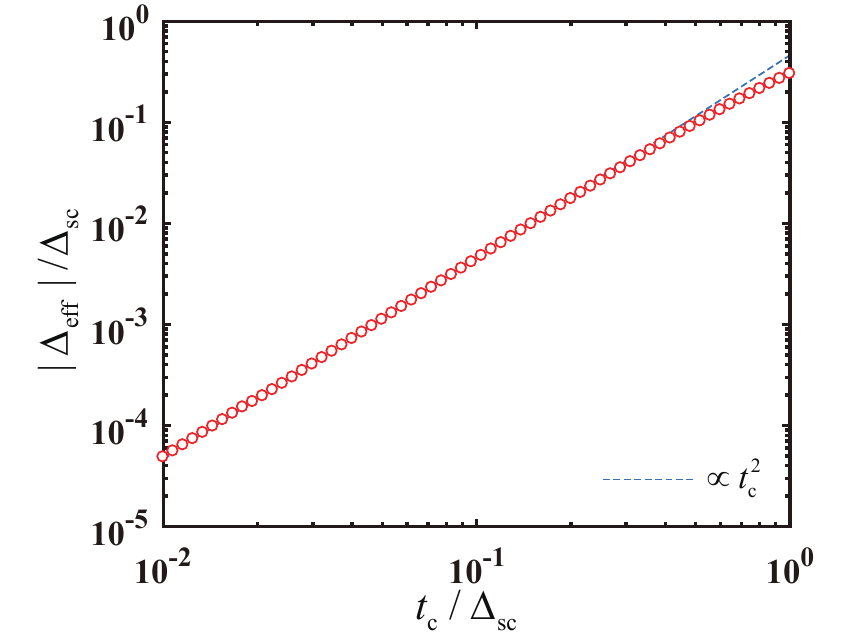}
\par\end{centering}
\caption{(color online) Effective pairing potential $\Delta_{\mathrm{eff}}$
  (empty circles) in the ferromagnet of a S/FM/S Josephson junction as a
  function of tunnel coupling $t_{\mathrm{c}}$ between the superconductor and
  ferromagnet. The blue dashed line depicts the magnitude of the
  proximity-induced pairing potential according to Eq. (12). Parameters used
  in the calculations are $N_{x}=60$, $\varphi=\pi/4$,
  $t_{\mathrm{sc}}=t_{\mathrm{fm}}=5\Delta_{\mathrm{sc}}$,
  $\lambda_{\mathrm{M}}=-0.7\Delta_{\mathrm{sc}}$,
  $\mu_{\mathrm{fm}}=-4.8\Delta_{\mathrm{sc}}$, and
  $\mu_{\mathrm{sc}}=-5\Delta_{\mathrm{sc}}$.}
\label{FIG_EffectivePairingPotential}
\end{figure}

\begin{figure*}
\centering
\includegraphics[width=0.32\textwidth]{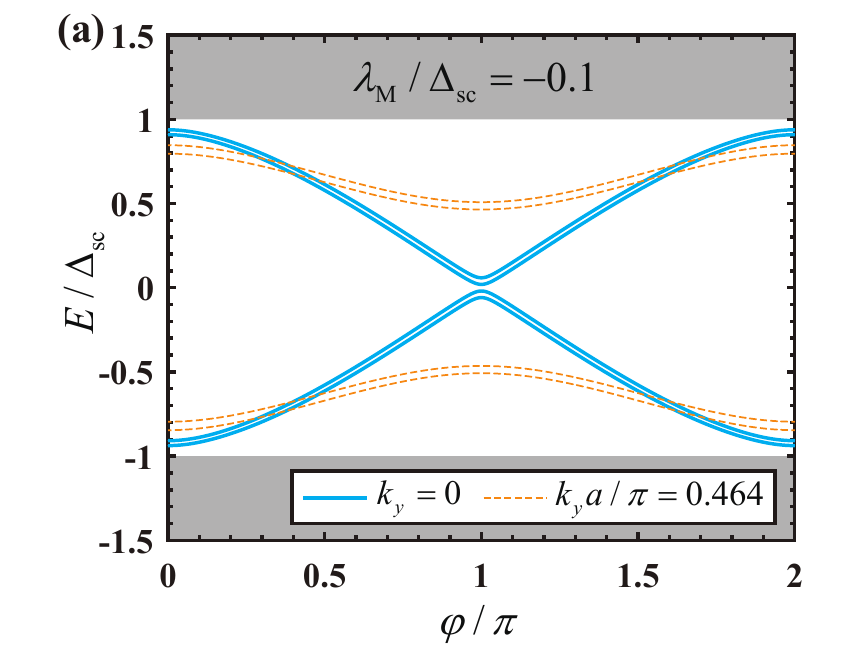}
\includegraphics[width=0.32\textwidth]{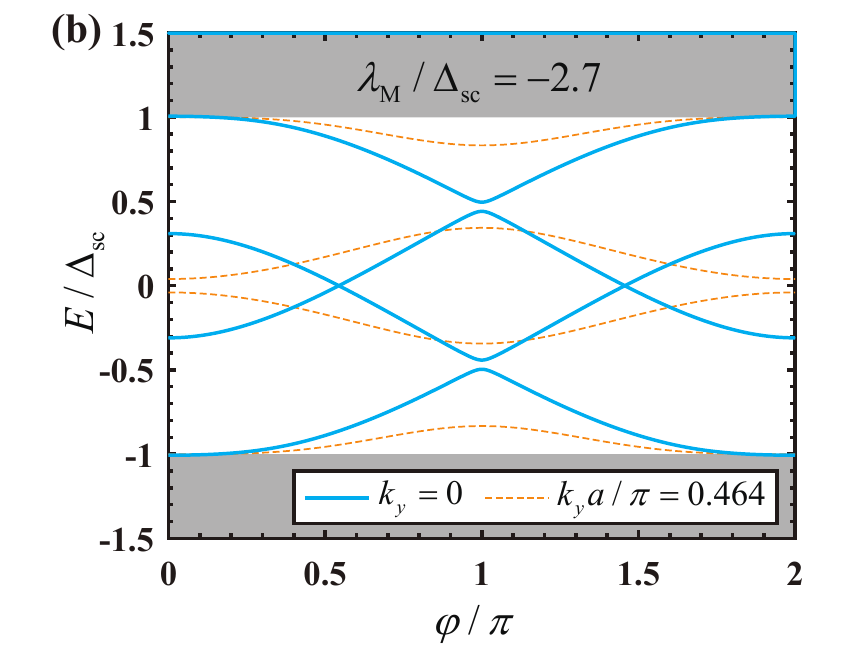}
\includegraphics[width=0.32\textwidth]{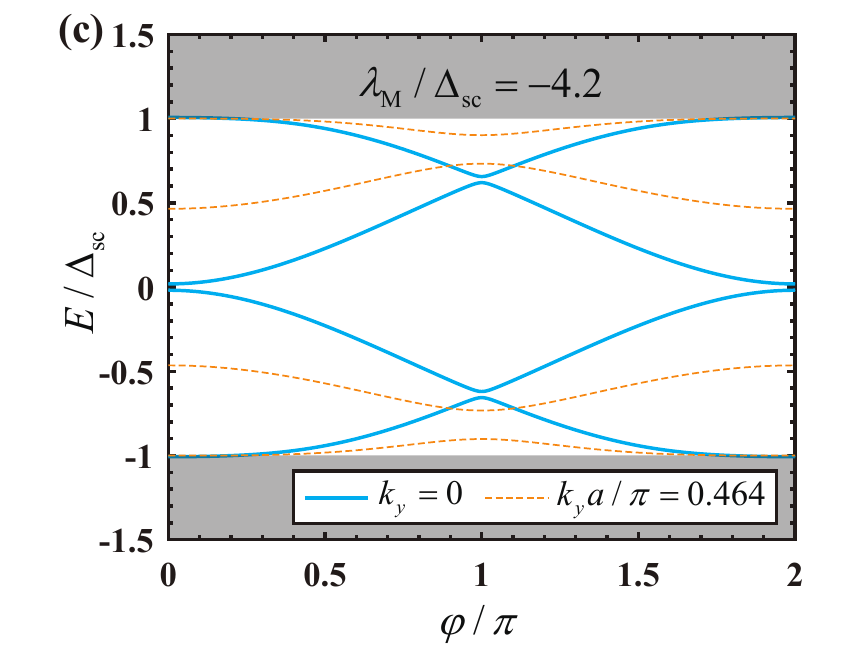}
\caption{(color online) Energy-phase relation of the subgap modes in an S/FM/S
  Josephson junction for (a) small
  ($\lambda_\mathrm{M}/\Delta_\mathrm{sc}=-0.1$), (b) intermediate
  ($\lambda_\mathrm{M}/\Delta_\mathrm{sc}=-2.7$), and (c) large
  ($\lambda_\mathrm{M}/\Delta_\mathrm{sc}=-4.2$) spin splitting.  The energies
  are plotted as a function of superconducting phase difference $\varphi$ at
  small ($k_y=0$, blue sold curves) and large ($k_y a/\pi=4.464$, orange
  dashed curves) momenta.  Other parameters used in the calculation:
  $N_{x}=60$, $\mu_{\mathrm{sc}}=-5\Delta_\mathrm{sc}$,
  $\mu_{\mathrm{fm}}=-5.2\Delta_\mathrm{sc}$, and
  $t_\mathrm{c} = t_{\mathrm{fm}}= t_{\mathrm{sc}} = 5 \Delta_\mathrm{sc}$.}
\label{FIG_ContinuumModelResults}
\end{figure*}

Representative examples of subgap modes are shown in Fig.~\ref{FIG_ThreeRegimes}. In general, the dispersion relation exhibits qualitatively different profiles depending on the characteristics of the isolated ferromagnet: metallic, half-metallic, and insulating.
When the ferromagnet is metallic, both spin bands have a Fermi surface and the coupling to the superconductor opens a proximity-induced superconducting gap at the Fermi level. The interplay between proximity effect and spin splitting has been the subject of intensive studies in various contexts, including spintronics applications,\cite{Buzdin05a} and we will not consider this case.
Instead, we will focus in this article on an insulating (upper panels of Fig.~\ref{FIG_ThreeRegimes})  or half-metallic\cite{Visani12a,Eschrig08a,Keizer06a,Bratkovsky97a} ferromagnet (lower panels). 
In a half-metallic ferromagnet, one of the spin bands is lifted above the Fermi level. One may think that, as superconducting pairing at the Fermi energy is impossible, proximity-induced subgap states are absent in this case. For an insulating ferromagnet, both spin bands are lifted above the Fermi level and, again, one may not expect subgap states. These arguments, however, are mostly based on the so-called semiconductor model, which incorporates only the energy gap around the Fermi level of the superconductor but does not take properly into account the superconducting pairing correlation. As shown in Ref.~\onlinecite{2018_PRB_Costa} for the insulating ferromagnet case, there are two different kinds of subgap bound states in the ferromagnetic region: the usual Andreev bound states and the Yu-Shiba-Rusinov states.

% Since the S/FM/S junction model possess translational invariance in
% the $y$-direction,
% different traveling modes along this direction
% are labeled in terms of the corresponding wave number $k_{y}$.
%%
% As shown in Fig.~\ref{FIG_ThreeRegimes} (a,d), the energies of the subgap
% states as a function of $k_{y}$ behave quite differently depending on the
% choice of $\lambda_{\mathrm{M}}$, $\mu_{\mathrm{sc}}$, and
% $\mu_{\mathrm{fm}}$, thus we classify the behaviors into two types: (i)
% Ferromagnet insulator (FI) shown in Fig.~\ref{FIG_ThreeRegimes}(a), energy
% dispersion from the isolated ferromagnet (the blue dashed curves in
% Fig.~\ref{FIG_ThreeRegimes}) does not cross the Fermi energy of the bulk
% SCs; (ii) Ferromagnet metal/half-metal shown in
% Fig.~\ref{FIG_ThreeRegimes}(d), if the energy dispersion crosses the Fermi
% energy.
%% 
% \footnote{By half-metal, we mean that only one spin component from the
% energy spectrum of the isolated ferromagnet crosses the Fermi energy of the
% bulk SC. This is usually achieved with a moderately larger value of
% $\lambda_{\mathrm{M}}$ in relation to
% $|\mu_{\mathrm{fm}}-\mu_{\mathrm{sc}}|$.}

In Fig.~\ref{FIG_ThreeRegimes}, the two panels on the left describe a bad superconductor-ferromagnet interface (i.e., a weak superconductor-ferromagnet coupling). As seen in panel (a), the dispersion for the S/FM/S Josephson junction with an insulating ferromagnet is relatively simple, as it is nearly quadratic in $k_y^2$. Instead, the dispersion for the junction with a half-metallic ferromagnet, shown in panel (d), is far richer even in this limit. In particular, the bare dispersion of the isolated half-metallic ferromagnet (dashed curves) is preserved near the Fermi level, but there is an anti-crossing of the dispersion curves belonging to different spins [see the circled region in Fig.~\ref{FIG_ThreeRegimes}(d)], which we attribute to proximity-induced superconductivity.\cite{Capogna96a,Volkov95a,McMillan68a} To confirm this interpretation, we have derived an effective model for the ferromagnet, valid at small $t_c$. In this regime, the low-energy properties of the system, especially the modes along the ferromagnetic wire below the superconducting gap $\Delta_\mathrm{sc}$, can be described by integrating out the superconducting electron operators, $\hatc_{x,y,\sigma}$ and $\hatc_{x,y,\sigma}^\dag$. Doing so (see Appendix~\ref{SFS2D::sec:A}), we obtain the effective model:
\begin{equation}
\label{SFS2D::eq:5}
\hatH_\mathrm{eff} = \hatH_\mathrm{fm}
+
\left(\Delta_\mathrm{eff} \sum_y\hatf_{y,\uparrow}^\dag\hatf_{y,\downarrow}^\dag
+ \mathrm{h.c.}\right) \,,
\end{equation}
where 
%\color{red}
\begin{equation}
\label{SFS2D::eq:4}
\Delta_{\mathrm{eff}} = \frac{t_\mathrm{c}^{2}}{4}
\left(\frac{1}{\Delta_\mathrm{L}^*} + \frac{1}{\Delta_\mathrm{R}^*}\right)
\end{equation}
%\color{black}
is the proximity-induced pairing potential inside the ferromagnet. As shown in Fig.~\ref{FIG_EffectivePairingPotential}, this estimate agrees well with the full numerical solution. In particular, the gap at the avoided crossing is proportional to $t_c^2$, the squared tunneling amplitude between the superconductor and ferromagnet, as expected in the weak-tunneling limit.\cite{Volkov95a}

If we now consider a stronger superconductor-ferromagnet coupling $t_c$, the difference between half-metallic and insulating ferromagnet in the S/FM/S Josephson junction gradually disappears. The nearly-transparent interface limit is illustrated by Figs.~\ref{FIG_ThreeRegimes}(b) and (e), and the evolution of the whole profiles as function of $t_c$ is plotted in Figs.~\ref{FIG_ThreeRegimes}(c) and (f). For both half-metallic and insulating ferromagnets, we note that in the limit of transparent interface the dispersion is close to the superconducting gap around $k_y \approx 0$. This behavior is consistent with previous studies in quasi-1D junctions.\cite{2018_PRB_Costa,2019_PRB_Rouco} On the other hand, the full dependence on $k_y$ shows that the dispersion is almost flat. An unusual feature, which was not expected from the treatment of quasi-1D junctions, is the appearance of local minima at large momenta ($k_y\approx\pm k_F$), seen in both panels (b) and (e) of Fig.~\ref{FIG_ThreeRegimes}.

To test the robustness of this behavior, we have also obtained the dispersion from the continuum model of Eq.~(\ref{SFS2D::eq:1}). Details of the solution are presented in Appendix~\ref{SFS2D::sec:B} and numerical results are shown in Fig.~\ref{FIG_ThreeRegimes_Cont}, which can be directly compared to Fig.~\ref{FIG_ThreeRegimes}. As seen, the overall features of the dispersion are in good agreement between the two models. In particular, in the limit of weak superconductor-ferromagnet coupling, see panels (a) and (d) of Fig.~\ref{FIG_ThreeRegimes_Cont}, we recover the bare dispersion of the isolated ferromagnet with anti-crossing points induced by proximity effect (in the half-metallic case). When the interface becomes more transparent, i.e., the height of the $\delta$-function barriers $l_0V_0$ decreases, the dispersion of the subgap states approaches $E \approx \pm \Delta_{\mathrm{sc}}$ and significantly flattens around $k_y=0$.

Despite the qualitative agreement, some detailed features differ. Especially, it is more difficult to obtain minima in the dispersion at finite $k_y$ based on the continuum model. Only very faint minima appear in Fig.~\ref{FIG_ThreeRegimes_Cont}(e) and the minima are absent in the insulating case, see Fig.~\ref{FIG_ThreeRegimes_Cont}(b). This behavior might be related to relatively small $k_y\sim 0.5 k_{F}$ at which the subgap states merge the superconducting bulk excitations in Fig.~\ref{FIG_ThreeRegimes_Cont}. Even if the dispersion at large $k_y$  is more sensitive to the details of the model, the occurrence of a nearly flat dispersion in a large range of $k_y$ values is robust feature of the strong coupling regime.

\section{$0$-$\pi$ transition}
\label{SFS2D::sec:5}

The competition of superconductivity with various correlation effects, such as strong electron-electron
interactions\cite{2004_PRB_Choi,2010_PRL_Zitko,Martin-Rodero11a} or ferromagnetism,\cite{2018_PRB_Costa,2019_PRB_Rouco,2006_RMP_Balatsky} has been known to drive a quantum phase transition which, in Josephson junctions, corresponds to the so-called 0-$\pi$ transition. 
In the zero phase, the ground state of the Josephson junction occurs with zero phase difference between the two superconductors, and the supercurrent is positive when $0<\varphi<\pi$. In the $\pi$ phase, the ground state occurs when the phase difference is $\pi$, and the supercurrent through the Josephson junction is negative.

% In previous studies, it has been shown that the S/FM/S junction can
% be either a $0$ or a $\pi$ junction depending on the strength of
% the magnetic spin splitting $\lambda_{\mathrm{M}}$ \cite{2018_PRB_Costa}.
In our system, a similar behavior occurs as the spin splitting in the ferromagnet is varied. As expected, at sufficiently small (large) values of $\lambda_\mathrm{M}$ the junction is in the 0 ($\pi$) phase. However, since the Josephson junction has two dimensions, a more detailed understanding of the $0$-$\pi$ transition should take into account the dependence on $k_y$. In an intermediate regime around the transition point we find that the character of the subgap states is mixed, i.e., some subgap states favor the zero phase, while the others favor the $\pi$ phase (depending on the wave number $k_y$). In other words, for a range of intermediate values of $\lambda_\mathrm{M}$ some of the subgap states in the zero ($\pi$) phase give a negative (positive) contribution to the current, which is opposite to the total Josephson current. Since the effect  depends on $k_y$, it could be probed by measuring the Josephson current as function of incident angle to the junction interface. 

\begin{figure}
\centering
\includegraphics[width=70mm]{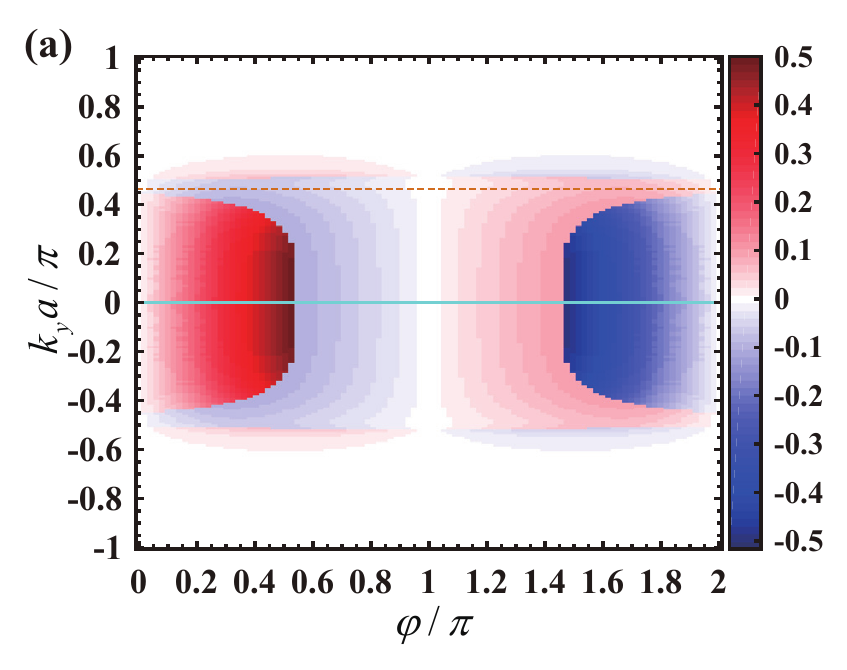}
\includegraphics[width=70mm]{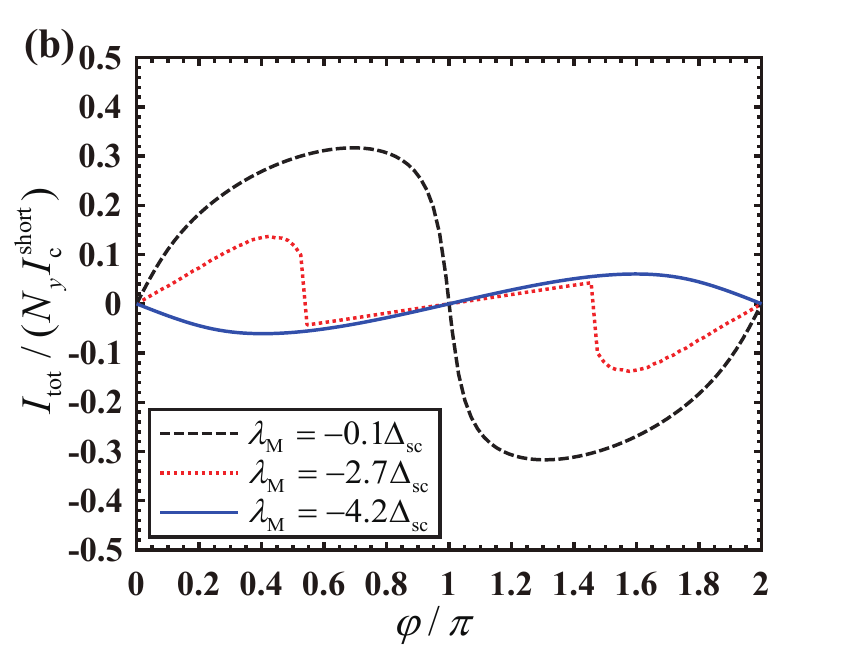}
\caption{(color online) Josephson current in various regimes. (a) Plot of the
  wave-vector resolved Josephson current for the intermediate regime of the
  previous Fig.~\ref{FIG_ContinuumModelResults}(b). The color scale refers to
  $I_{k_{y}}(\varphi)/I_{\mathrm{c}}^{\mathrm{short}}$, where
  $I_{\mathrm{c}}^{\mathrm{short}}=|e|\Delta_{\mathrm{sc}}/\hbar$ and
  $I_{k_{y}}(\varphi)$ is computed from Eq.~(\ref{Iky}).  The horizontal lines
  mark the two wavevectors of Fig.~\ref{FIG_ContinuumModelResults}(b).
  (b) Total Josephson current as a function of the superconducting phase
  difference. The black dashed, red dotted, and blue solid curves are plotted
  with the same parameters of panels (a), (b), and (c) of
  Fig.~\ref{FIG_ContinuumModelResults}, respectively.}
\label{FIG_Jcurrent}
\end{figure}

The behavior described above is illustrated by
Fig.~\ref{FIG_ContinuumModelResults}, where the energies of the subgap
states as a function of $\varphi$ are shown, for two different values of $k_y$.
%, calculated numerically based on the tight-binding model in Eq.~\eqref{Hamiltonian}. 
We observe that for a small spin splitting [Fig.~\ref{FIG_ContinuumModelResults}(a)] the  $k_y=0$ and large-$k_y$ phase dependencies both favor a Josephson junction in the zero phase, with the large-$k_y$ subgap state (orange dashed curves) giving a smaller contribution to the current. At  intermediate spin splitting [Fig.~\ref{FIG_ContinuumModelResults}(b)] the large-$k_y$ quasiparticles are in a regime favoring the $\pi$ phase. However, the phase dependence at $k_y=0$ is qualitatively different, and it is not obvious if these states favor the zero phase or not. To clarify the issue we compute explicitly the wavevector-resolved Josephson current:
\begin{equation}\label{Iky}
I_{k_{y}}(\varphi)=-\frac{|e|}{\hbar}\sum_{n=1}^2 \frac{\partial E_{n}(k_{y})}{\partial\varphi},
\end{equation}
where the summation over $n$ is restricted to sub-gap states with positive energy $E_{n}$. In Eq.~(\ref{Iky}) we have assumed the zero temperature limit. For the $k_y=0$ subgap states of Fig.~\ref{FIG_ContinuumModelResults}(b), Eq.~(\ref{Iky}) gives a positive current around $\varphi \simeq 0$, which changes sign when approaching $\varphi = \pi$. The full dependence of $I_{k_{y}}(\varphi)$ is presented in Fig.~\ref{FIG_Jcurrent}(a), showing the nontrivial behavior of the sign in the `mixed' regime. Finally, as the spin splitting increases further [Fig.~\ref{FIG_ContinuumModelResults}(c)], the subgap states favor the $\pi$ phase over the entire range of $k_y$. 

For the three panels of Fig.~\ref{FIG_ContinuumModelResults}, we have also obtained in Fig.~\ref{FIG_Jcurrent}(b) the total Josephson current, computed from Eq.~(\ref{Iky}) as $I_\mathrm{tot}(\varphi)=\sum_{k_y} I_{k_{y}}(\varphi) $. Due to the periodic boundary condition along $y$, the wavevector is discretized as $k_y=2\pi j/N_ya$ ($j=1,2,\ldots N_y$) where $N_y a$ is the total width of the junction (see Fig.~\ref{FIG_junction_schematics}). The phase dependence in Fig.~\ref{FIG_Jcurrent}(b) is consistent with previous discussions. While the smallest (largest) value of $\lambda_\mathrm{M}$ leads to a zero ($\pi$) junction, at intermediate spin splitting both zero and $\pi$ states are locally stable to small variations of the phase. A discontinuous behavior of $I_\mathrm{tot}(\varphi)$ (red dotted curve) is actually typical of the 0-$\pi$ transition region. For example, a similar dependence on $\varphi$ is obtained considering the supercurrent through a quantum dot in a regime of intermediate couplings.\cite{2004_PRB_Choi} The strong deviations from the sinusoidal dependence at small $\lambda_M$ can also be understood qualitatively form a comparison to the quantum dot model.
Since the result in Fig.~\ref{FIG_Jcurrent} is computed with a transparent barrier, $t_\mathrm{c}=t_\mathrm{sc}$, the small-$\lambda_\mathrm{M}$ limit bears similarity to the supercurrent through a quantum dot in the limit of resonant transmission.
In that case, a Josephson current with the asymmetric phase dependence $I_\mathrm{c}^\mathrm{short} \sin\varphi/2$ (for $-\pi<\varphi <\pi$) is approached, where $I_\mathrm{c}^\mathrm{short} = |e|\Delta_\mathrm{sc}/\hbar$ is the maximum current for a single 1D channel.\cite{Furusaki91a, 1991_PRL_Beenakker, 2004_PRB_Choi}
In our system, we see in Fig.~\ref{FIG_ContinuumModelResults}(a) that the $k_y=0$ states follow closely this ideal limit but the large-$k_y$ states largely depart from it.
As a result of these large-$k_y$ contributions, the \emph{total} current does not reach the maximum allowed value, $N_y I_\mathrm{c}^\mathrm{short}$, and deviates from the `period-doubling' dependence $\propto \sin\varphi/2$ (which would be discontinuous at $\varphi=\pi$), see the black dashed curve of Fig.~\ref{FIG_Jcurrent}(b).   

Finally, we stress that, although Fig.~\ref{FIG_ContinuumModelResults} was computed with the tight-binding model of Eq.~\eqref{Hamiltonian}, the same behavior is found from the continuum model. Explicit calculations based on  Eq.~\eqref{BdG_Hamiltonian} are presented in Appendix~\ref{SFS2D::sec:B}, where the energy-phase relation of the subgap modes is shown in Fig.~\ref{FIG_Phase_Cont}. 

% The energy-phase relation shown in Fig.~\ref{FIG_ContinuumModelResults}
% further illustrates the important role of $k_{y}$ in establishing the
% $0$-$\pi$ transition. In previous study \cite{2018_PRB_Costa}, the S/FM/S
% junction was assumed to be effective one-dimensional along the $x$-direction,
% thus the effect of the transverse wave vector $k_{y}$ is not quite
% obvious. However, the results here show that energy-phase relation can be
% modified as $k_{y}$ increasing. This situation is most clearly seen in
% Fig.~\ref{FIG_ContinuumModelResults}(b): As $\lambda_{\mathrm{M}}$ kept
% increasing the bound states start to display the complete feature of a $\pi$
% junction firstly for states that are more parallel to the junction interface,
% i.e., with a larger value of $k_{y}$. Therefore, if properly taken into
% account, the transverse wave vector may have a positive effect on the
% $0$-$\pi$ transition, e.g., inducing the transition to a $\pi$-junction with a
% smaller requirement on the strength $\lambda_{\mathrm{M}}$.

\section{Quasi-Particle Flat-Band Effects}
\label{SFS2D::sec:4}

As discussed in Sec.~\ref{SFS2D::sec:3}, the regime of large $t_c$ leads to subgap states with a flat dispersion. This feature might be useful to engineer the wave functions of subgap modes along an inhomogeneous ferromagnetic junction. In general, transport properties become extraordinarily sensitive to defects or impurities when the effective mass is extremely heavy. One possibility here is to introduce domain walls along the ferromagnet, so as to cause strong scattering at those interfaces. Note that recent developments of the state-of-art technologies in spintronics \cite{Caretta20a, Zutick04a, Wolf01a} allow one to control such domain walls with high precision and speed.

%%%%%%%%%%%%%%%%%%%%%%%%%%%%%%%%%%%%%%%%%%%%%%
\begin{figure}
\begin{centering}
\includegraphics[width=0.38\textwidth]{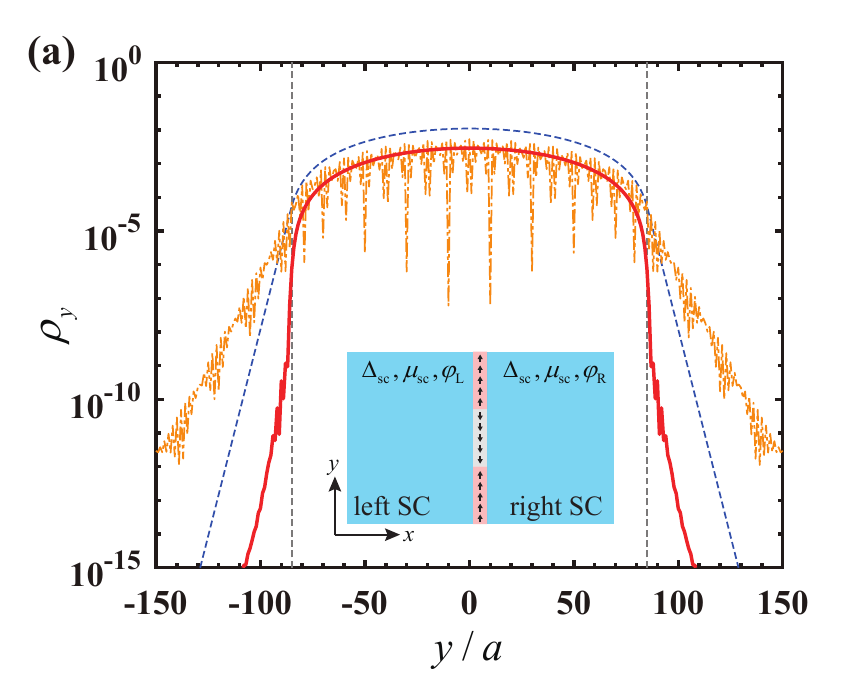}
\includegraphics[width=0.38\textwidth]{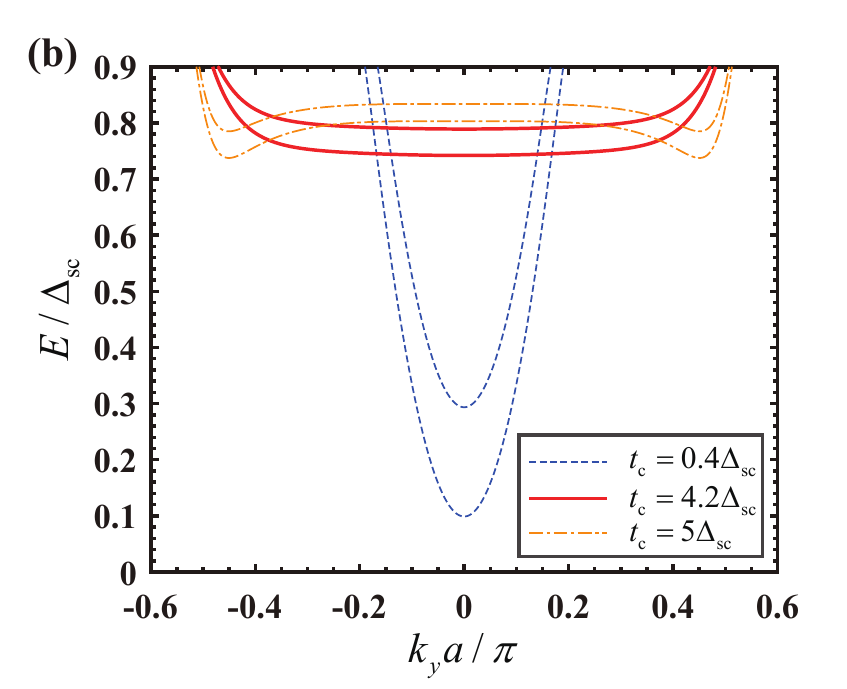}
\par\end{centering}
\caption{(color online) (a) Ground state density profile $\rho_{y}$ for three
  representative values of $t_{\mathrm{c}}= 0.4 \Delta_\mathrm{sc}$ (blue dashed
  curve), $4.2\Delta_\mathrm{sc}$ (thick red solid curve), and $5\Delta_\mathrm{sc}$
  (orange dashed curve). Black vertical dashed lines indicate the domain walls
  at $y=\pm 85a $ (see also the inset). (b) The corresponding subgap energy
  dispersions for a uniform domain. Parameters used in the calculations:
  $N_{x}=60$, $N_{y}=301$, $\varphi=\pi/4$,
  $\mu_{\mathrm{sc}}=\mu_{\mathrm{fm}}=5\Delta_{\mathrm{sc}}$,
  $t_{\mathrm{sc}}=t_{\mathrm{fm}}=5\Delta_{\mathrm{sc}}$,
  $\lambda_{\mathrm{M},1}=-0.1\Delta_{\mathrm{sc}}$, and
  $\lambda_{\mathrm{M},2}=0.1\Delta_{\mathrm{sc}}$.}
\label{FIG_WaveFunction}
\end{figure}
%%%%%%%%%%%%%%%%%%%%%%%%%%%%%%%%%%%%%%%%%%%%%%

In the following, we focus on an insulating ferromagnet and consider S/FM/S Josephson junctions with two domains. For the inhomogenous junction of Fig.~\ref{FIG_WaveFunction}, the inner domain has a positive spin splitting $\lambda_{\mathrm{M},1}>0$ while the outer domain (recall the periodic boundary condition in the $y$-direction) has a negative value $\lambda_{\mathrm{M},2}<0$.
For such inhomogeneous system, a quasiparticle of energy $\varepsilon$ is not a plane wave but has the generic form $\hat\gamma_\varepsilon = \sum_{y,\sigma} (u_{y,\sigma }\hat{f}_{y,\sigma}+v_{y,\sigma }\hat{f}^\dag_{y,\sigma}) + \hat{\gamma}_\mathrm{sc}$, where  $\hat{\gamma}_\mathrm{sc}$ is the contribution from the superconducting leads.
By considering the lowest-energy subgap state (with $\varepsilon>0$), we characterize the density profile along the ferromagnetic chain through
\begin{equation}
\rho_y = \sum_\sigma \left(|u_{y,\sigma}|^2 +|v_{y,\sigma}|^2  \right),
\end{equation}
which is plotted in Fig.~\ref{FIG_WaveFunction}(a) for three choices of $t_c$. As seen, the wave function profile changes abruptly across the domain walls. In the inner domain the wave function has the characteristics of a traveling wave, whereas it is evanescent in the other domain. The almost total reflection of the wave function incident from the inner to outer domain is significantly enhanced by the heavy effective mass, making the penetration depth around $t_c  = 4.2 \Delta_\mathrm{sc}$ (thick red curve) extremely short. For reference, we show in Fig.~\ref{FIG_WaveFunction}(b) the corresponding dispersion of subgap modes for the uniform S/FM/S junctions (i.e., assuming a constant  $\lambda_\mathrm{M}=\lambda_{\mathrm{M},1}>0$). The dispersions correspond well to the behavior of panel (a). For the smallest value of $t_c =0.4\Delta_\mathrm{sc}$ the flattening of dispersions has not occurred yet, thus the reflection of the wave functions at the domain walls is not so dramatic as for $t_c = 4.2 \Delta_\mathrm{sc} $. By further increasing $t_c$, the dispersion develops two minima at finite $k_y$, thus the quasiparticles are actually characterized by a relatively small effective mass, in agreement with the weaker reflection at the interface. At $t_c = 5 \Delta_\mathrm{sc} $, the fast oscillations seen in $\rho_y$ reflect the large wavevector difference between the two valleys. 

Despite the reasonable agreement between panels (a) and (b) of Fig.~\ref{FIG_WaveFunction}, the effective mass is not the only factor which determines the penetration length. For example, the relatively long penetration depth at $t_c = 5 \Delta_\mathrm{sc} $, see Fig.~\ref{FIG_WaveFunction}(a), seems difficult to explain only based on the effective mass. Another important factor to take into account should be the effective barrier at the domain wall. We note that in panel (b) the energy difference between the two $t_c = 5 \Delta_\mathrm{sc}$ subgap bands is significantly smaller than the splitting at $t_c = 0.4\Delta_\mathrm{sc}$. The smaller energy spitting might be related to a reduced potential step, seen by the quasiparticles when entering the opposite ferromagnetic domain. This effect would help explaining the longer penetration depth at $t_c = 5 \Delta_\mathrm{sc}$. In general, while the interface scattering between domains is a rather involved problem, Fig.~\ref{FIG_WaveFunction} indicates that the localization properties can be enhanced in the regime of good S/FM interfaces.

\begin{figure}
\begin{centering}
\includegraphics[width=0.38\textwidth]{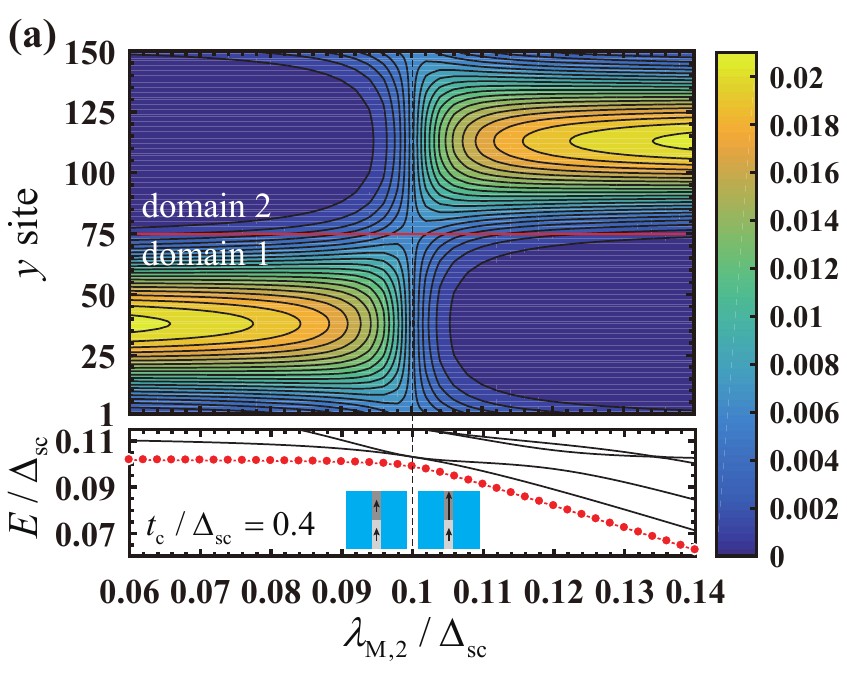}
\includegraphics[width=0.38\textwidth]{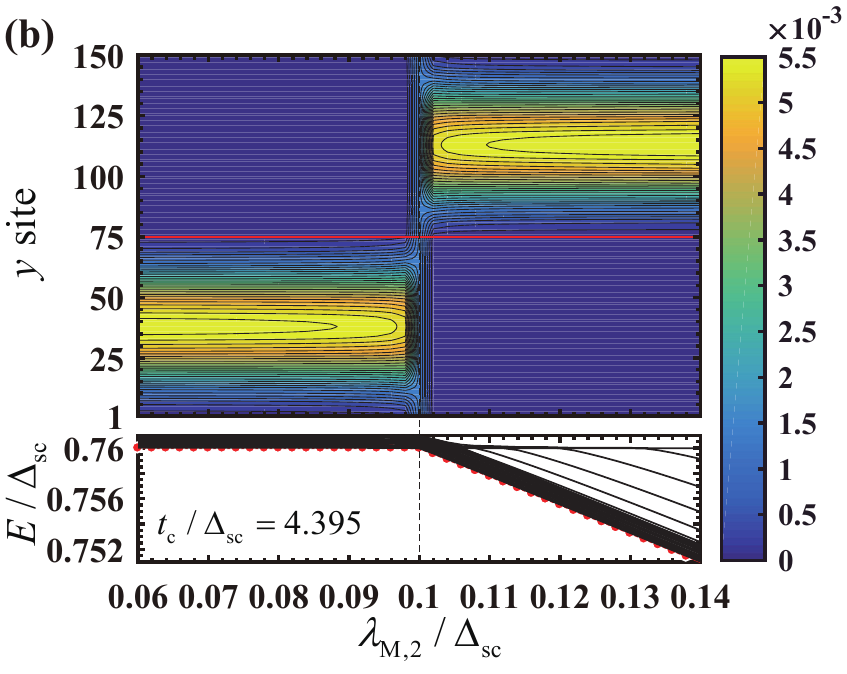}
\par\end{centering}
\caption{(color online) Switching behavior with parallel domains. In (a) we
  plot $\rho_y$ as a function of the spin splitting $\lambda_{\mathrm{M},2}$
  of the second ferromagnetic domain. The red horizontal line marks the domain
  wall and the black dashed vertical line corresponds to
  $\lambda_{\mathrm{M},2}=\lambda_{\mathrm{M},1}$ (homogeneous junction). As
  shown in the lower-panel inset, domain 2 spans from site $N_y/2+1$ to site
  $N_y$ (calculations were performed on a lattice with size
  $[-N_x,N_x]\times[1,N_y]$, where $N_{x}=60$ and $N_{y}=150$).  Panel (b) is
  the same of (a), except that the $t_c$ is chosen to yield a nearly `flat'
  quasiparticle dispersion. The two values of $t_\mathrm{c}$ are
  $0.4\Delta_{\mathrm{sc}}$ and $4.395\Delta_{\mathrm{sc}}$ for panel (a) and
  (b), respectively. Other parameters: $\varphi=\pi/4$,
  $\mu_{\mathrm{sc}}=-5\Delta_{\mathrm{sc}}$,
  $t_{\mathrm{sc}}=t_{\mathrm{fm}}=5\Delta_{\mathrm{sc}}$,
  $\mu_{\mathrm{fm}}=-5.2\Delta_{\mathrm{sc}}$, and
  $\lambda_{\mathrm{M},1}=0.1\Delta_{\mathrm{sc}}$.}
\label{FIG_DomainSwtich}
\end{figure}

Figure~\ref{FIG_DomainSwtich} further demonstrates effects of the flat bands on the population density of quasi-particles. Here, at variance with Fig.~\ref{FIG_WaveFunction}, the two domains have parallel spin polarization (both $\lambda_{\mathrm{M},1}$ and $\lambda_{\mathrm{M},2}$ are positive).
In panel (a) we plot $\rho_y$ as a function of the spin splitting $\lambda_{\mathrm{M},2}$ of one domain (with the spin splitting $\lambda_{\mathrm{M},1}$ fixed for convenience). As the spin splitting is varied, the population densities in the two domains are switched abruptly.
In Fig.~\ref{FIG_DomainSwtich}(b) we show that this effect is greatly enhanced when the tunnel coupling between the superconductor and ferromagnet is tuned to regime of flat bands, making the switching behavior much sharper.
One possible application of such a behavior is realizing a local switch of Josephson current through the modulation of the magnetization of ferromagnetic domains.

% Two-dimensional S/FM/S junction with multiple ferromagnetic domains along the
% $y$-direction may have potential applications. One such possibility is to
% switch the bound state wave function by locally tuning the magnetic splitting
% $\lambda_{\mathrm{M},1}$ or $\lambda_{\mathrm{M},2}$.  As shown in
% Fig.~\ref{FIG_DomainSwtich}, such manipulation could lead to a switching of
% wave functions which can be understood from the effective one-dimensional
% model Eq.~\eqref{effective_Hamiltonian} as the spin-spliting provides local
% spin-dependent potential wells. This behavior may have application in locally
% moving the quasi-particle that bounded at the ferromagnetic layer from one
% domain to the other, as suggested from the wave function behavior in
% Fig.~\ref{FIG_DomainSwtich}, and leads to local manipulation of the Josephson
% current through the S/FM/S junction. Furthermore, as shown in
% Fig.~\ref{FIG_DomainSwtich}(b) the switch of the wave function depends more
% sensitively on $\lambda_{\mathrm{M},1}/\lambda_{\mathrm{M},2}$ if the bound
% state dispersion is tuned to be `flat'.

\begin{figure*}
\begin{centering}
\begin{tabular}{ccc}
\includegraphics[width=55mm]{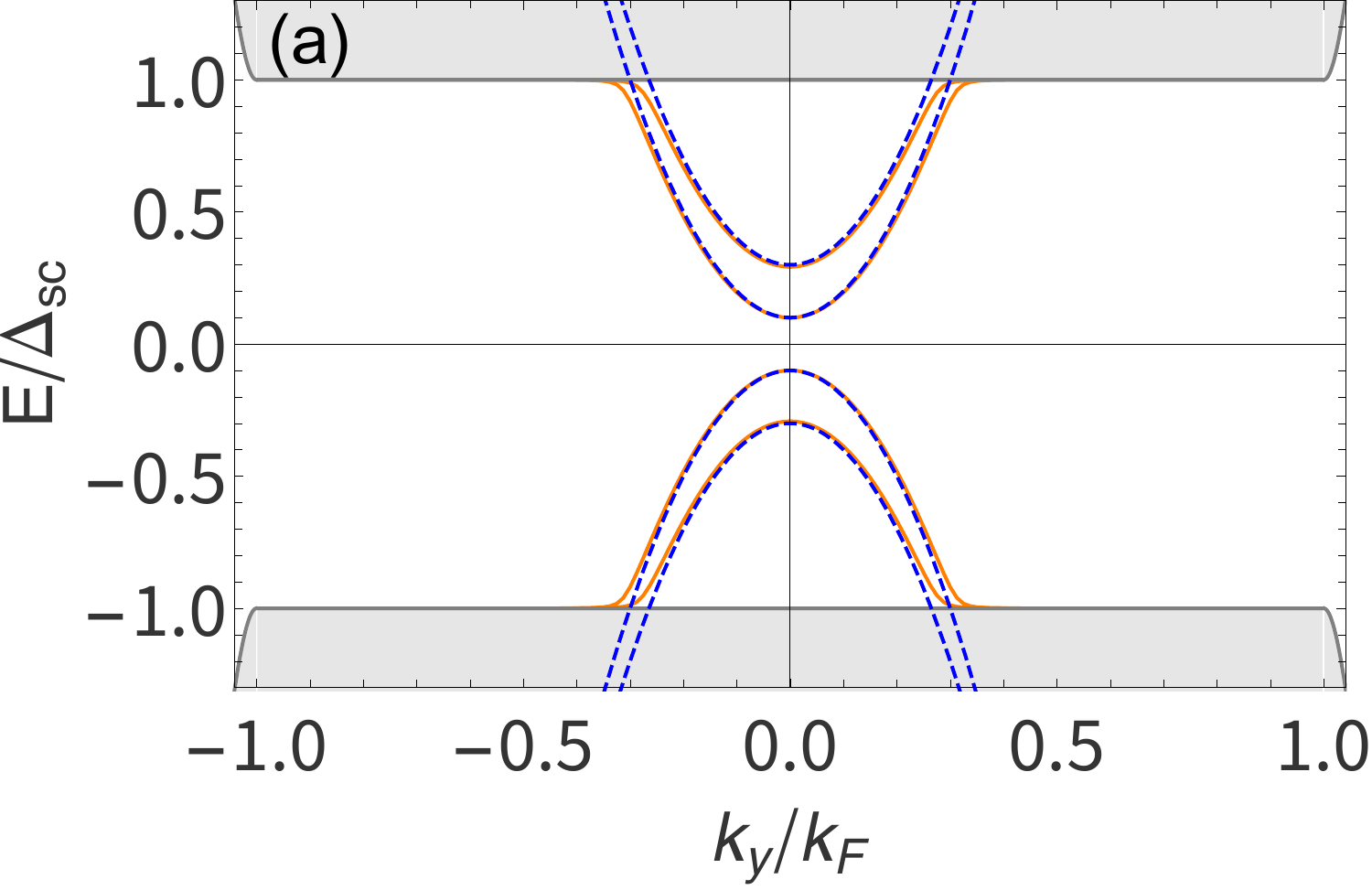}  &
\includegraphics[width=55mm]{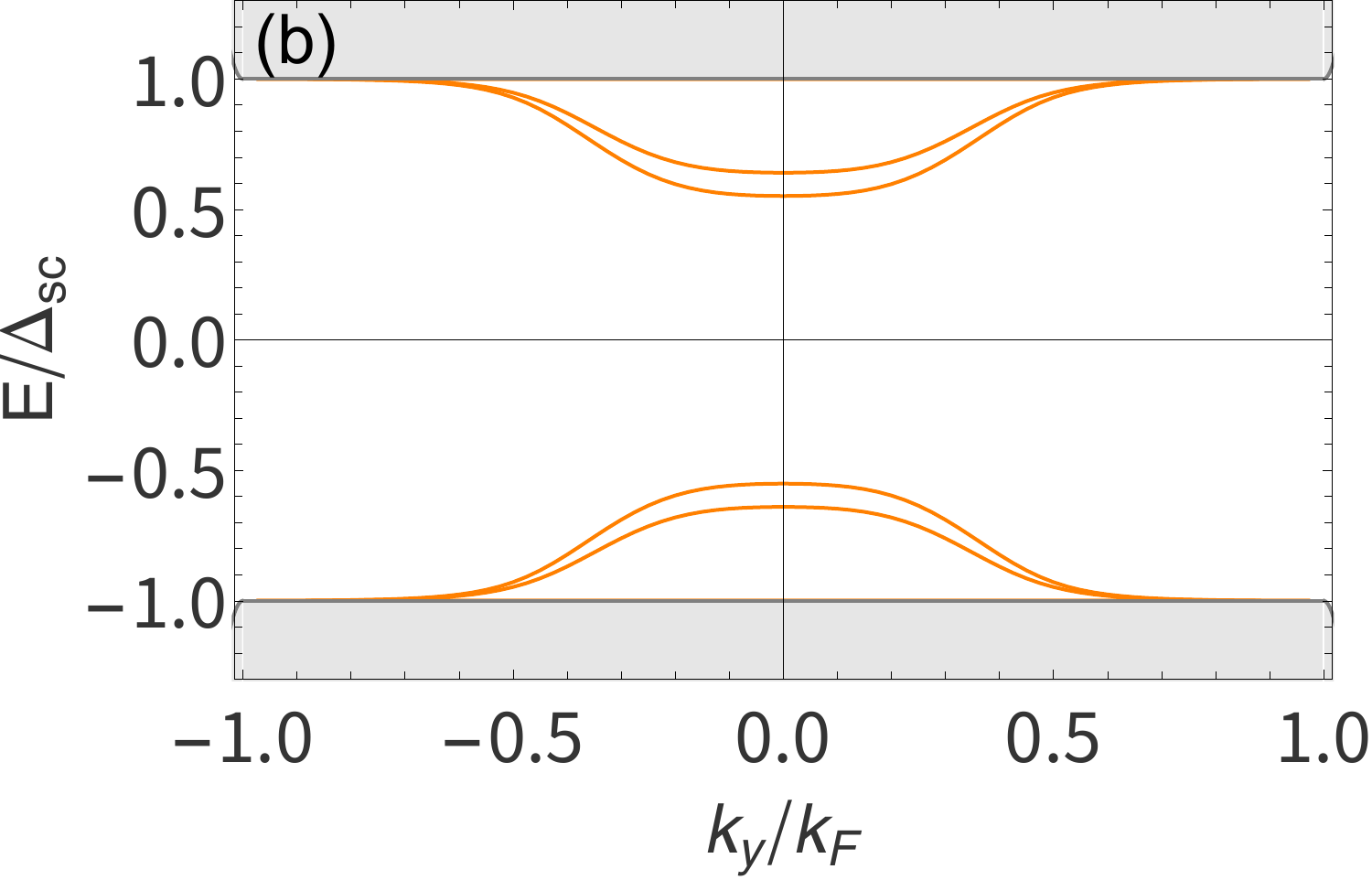}  &
\includegraphics[width=60mm]{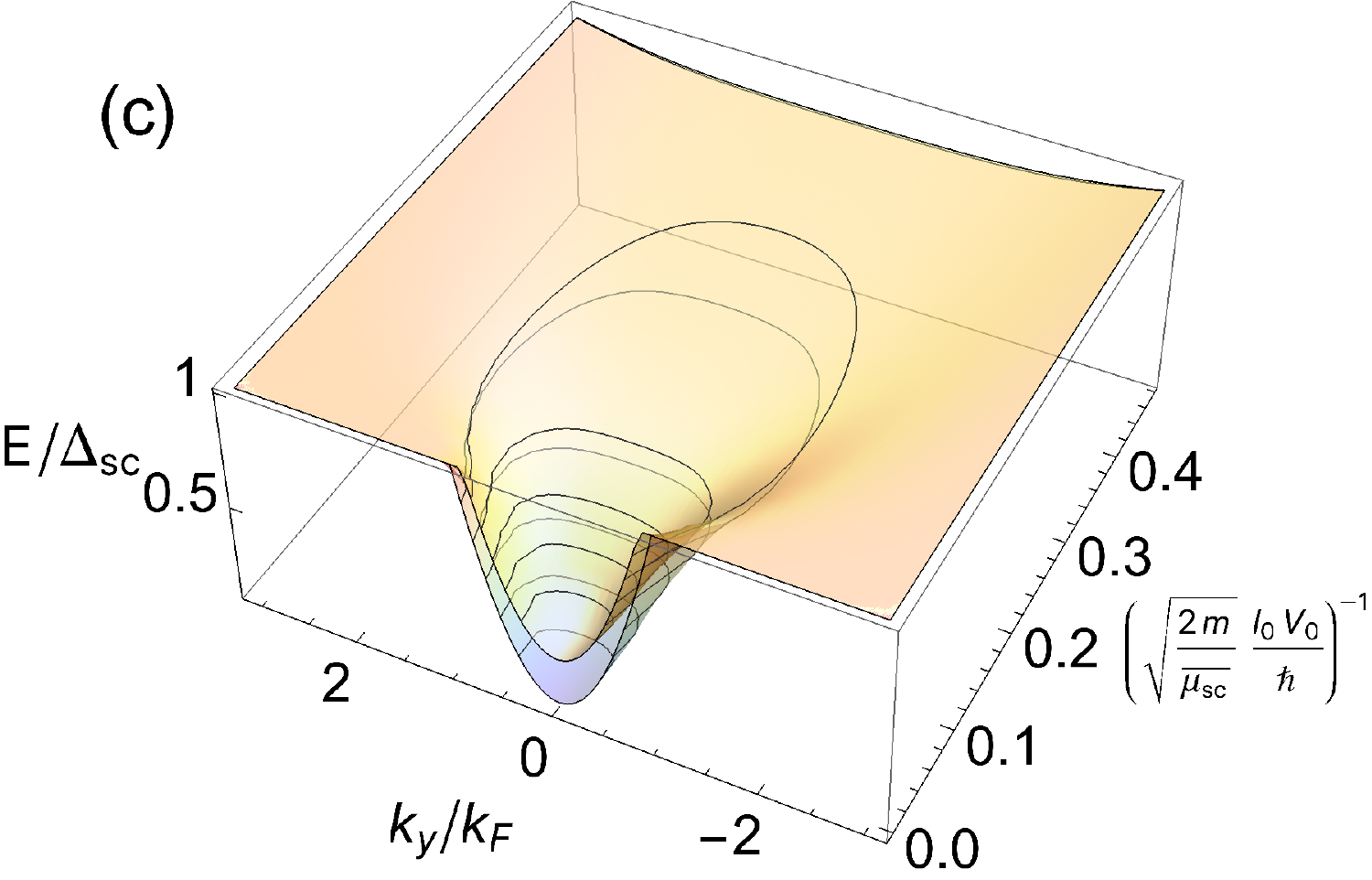}  \\
\includegraphics[width=55mm]{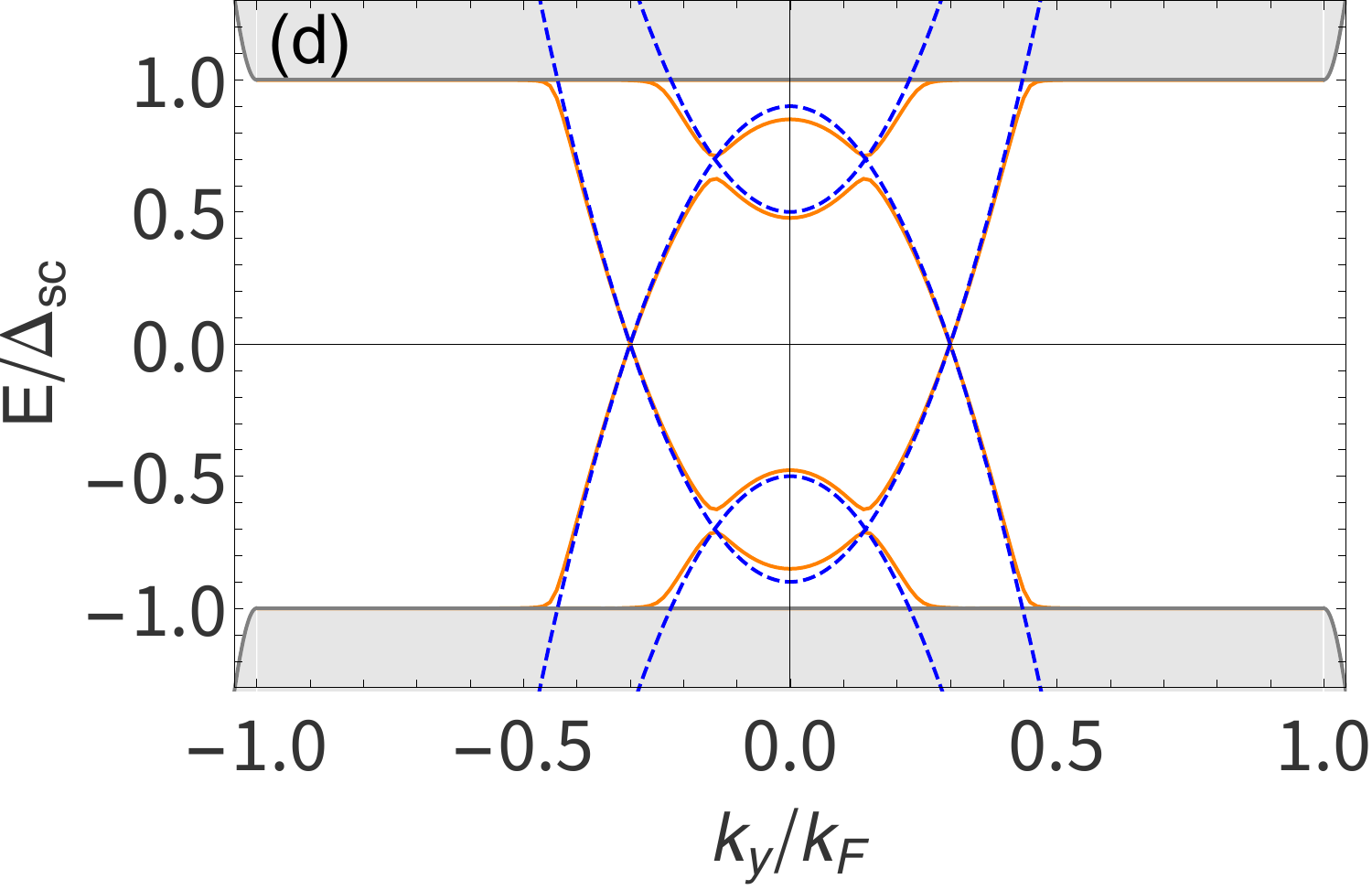}  &
\includegraphics[width=55mm]{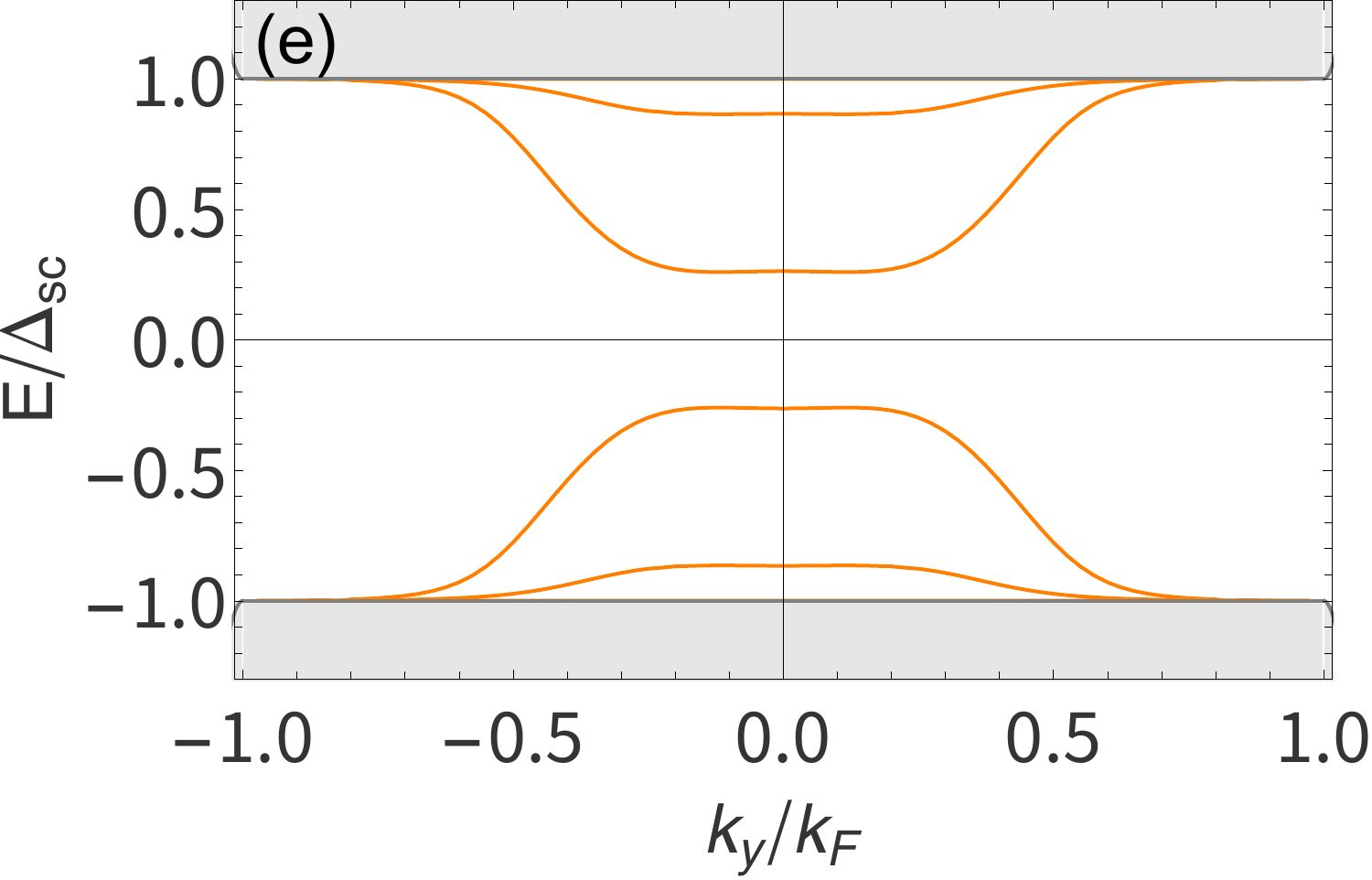}  &
\includegraphics[width=60mm]{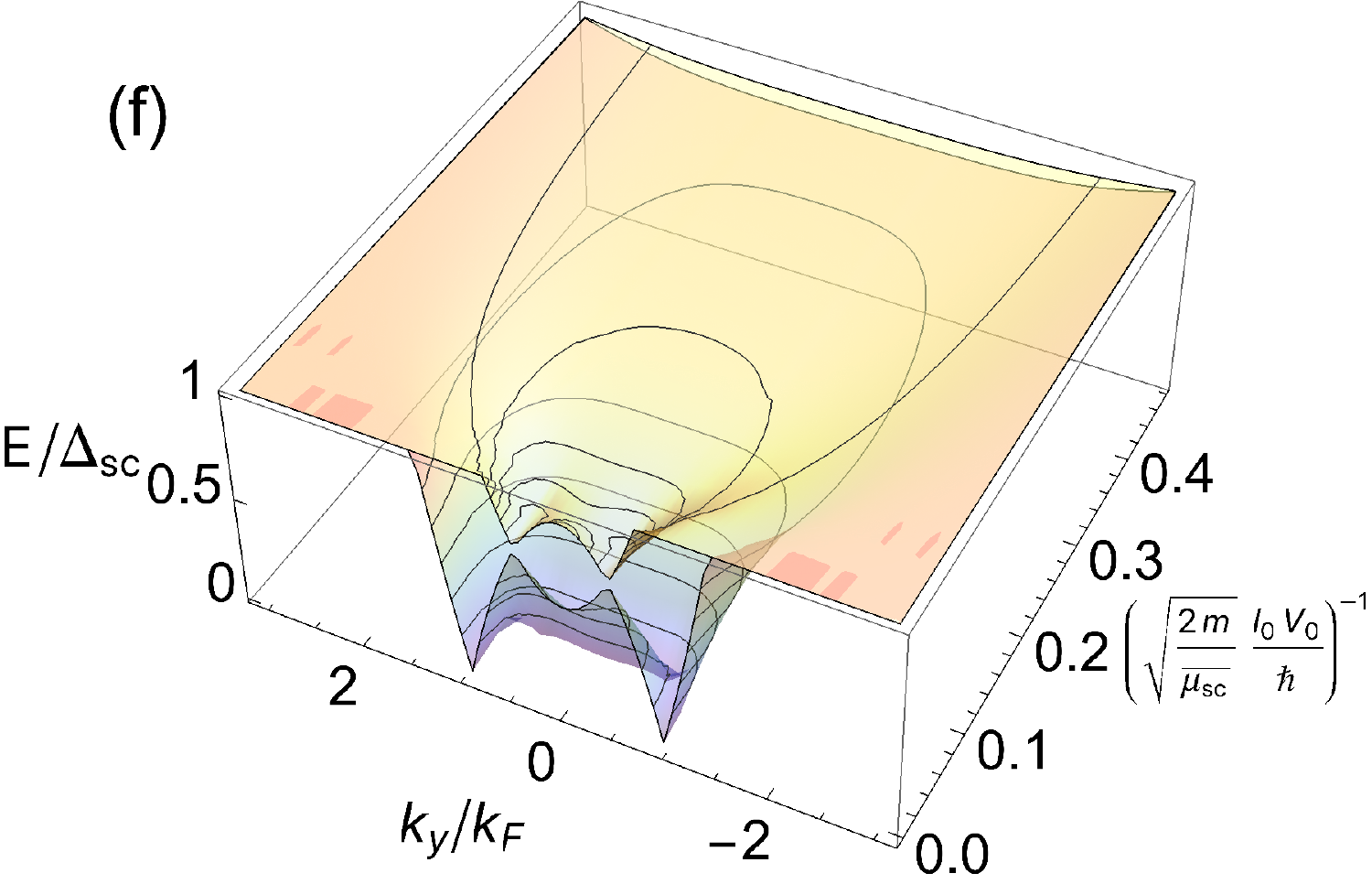}
\end{tabular}
\par\end{centering}
\caption{(color online) Energy spectrum of the S/F/S junction, obtained with
  the continuum model of Eq.~(\ref{SFS2D::eq:1}).  The transverse wave-vector
  $k_y$ is expressed in units of $k_F =
  \sqrt{2m\bar\mu_{\mathrm{sc}}}/\hbar$. In the upper panels we have used
  \color{blue} $\bar{\mu}^\mathrm{eff}_\mathrm{fm}=-0.2 \Delta_{\mathrm{sc}}$
  \color{black} and $\lambda_\mathrm{M}=0.1\Delta_{\mathrm{sc}}$. In the lower
  panels \color{blue}
  $\bar{\mu}^\mathrm{eff}_\mathrm{fm}=0.2 \Delta_{\mathrm{sc}}$ \color{black} and
  $\lambda_\mathrm{M}=0.7\Delta_{\mathrm{sc}}$. $\bar{\mu}_\mathrm{fm}$ is
  computed from $\bar{\mu}^\mathrm{eff}_\mathrm{fm}$ as described in the last
  part of Appendix~\ref{SFS2D::sec:B}. Other parameters:
  $\bar{\mu}_\mathrm{sc}=10 \Delta_{\mathrm{sc}}$, $L = k_F^{-1}$, while the
  inverse barrier height is given by \color{blue}
  $(l_0V_0)^{-1}= 0.01k_F/\bar\mu_\mathrm{sc}$ for panels (a) and (d) and
  $(l_0V_0)^{-1} = 0.075k_F/\bar{\mu}_\mathrm{sc}$ in panels (b) and (e).}
 \label{FIG_ThreeRegimes_Cont}
\end{figure*}

\section{Conclusion}
\label{SFS2D::sec:6}

We studied two-dimensional S/FM/S junctions, assuming that the narrow ferromagnetic strip is metallic, half-metallic, or insulating.
By investigating the subgap modes along the junction interface, we find that
they inherit the characteristics of the Yu-Shiba-Rusinov states
that originate from the interplay between superconductivity and ferromagnetism
in the magnetic junction. Such characteristics lead to several intriguing properties that are not observable in 1D or quasi-1D magnetic junctions:
First, the dispersion relation of the subgap modes shows characteristic
profiles depending on the transport state (metallic, half-metallic, or insulating) of
the ferromagnet, as well as the quality of the superconductor-ferromagnet
interface.
Second, the subgap modes induce a $0$-$\pi$ transition in the Josephson current across the junction as the spin splitting in the ferromagnet is increased.
The Josephson current density as a function of superconducting phase difference changes sharply with the momentum along the junction interface (i.e., the direction of the incident current).
Finally, for clean superconductor-ferromagnet interfaces (i.e., strong coupling between superconductors and ferromagnet), the subgap modes develop flat quasi-particle  bands. Such flat bands turn out to be useful in engineering the wave function of subgap modes along an inhomogeneous magnetic junction.
We note that the recent development of state-of-art technologies in spintronics \cite{Caretta20a, Zutick04a, Wolf01a} enables controlling domain walls in a ferromagnetic nanowire with high precision and speed, making our findings relevant for experiments.

\begin{acknowledgments}
Y.F. acknowledges support from NSFC (Grant No. 12005011). S.C. acknowledges support from the National Science Association Funds (Grant No. U1930402) and NSFC (Grants No. 11974040 and No. 12150610464). M.-S.C. was supported by the Ministry of Science and ICT of Korea (Grant Nos. 2017R1E1A1A03070681 and 2022M3H3A1063074) and by the Ministry of Education of Korea through the BK21 program.
\end{acknowledgments}

\appendix

\begin{figure*}
\begin{centering}
\begin{tabular}{ccc}
\includegraphics[width=18cm]{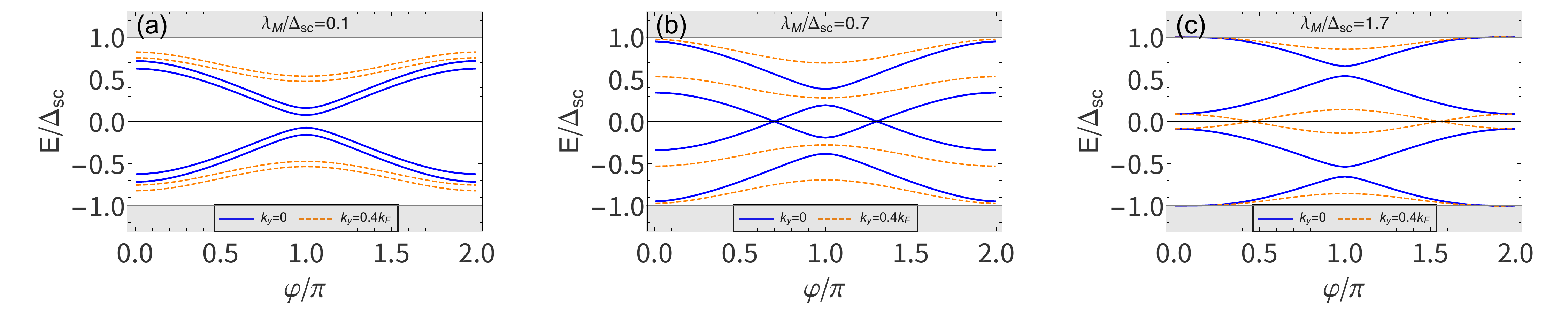}
\end{tabular}
\par\end{centering}
\caption{(color online) Energy-phase relation for the subgap states at
  $k_y = 0$ (blue solid curves) and $k_y = 0.4 k_F$ (orange dashed curves),
  obtained by using the continuum model in Eq.~(\ref{SFS2D::eq:1}). Panels
  (a), (b), and (c) are for
  $\lambda_\mathrm{M} = 0.1\Delta_{\mathrm{sc}}, 0.7\Delta_{\mathrm{sc}}$, and
  $ 1.7\Delta_{\mathrm{sc}}$, respectively. Other parameters:
  $\bar{\mu}_\mathrm{sc} = 10\Delta_{\mathrm{sc}}$, 
  $\bar{\mu}^\mathrm{eff}_\mathrm{fm}=0.4\Delta_{\mathrm{sc}}$,
  $L = k_F^{-1}$, and
  $\left( l_0V_0 \right)^{-1} =0.075 k_F/\bar{\mu}_\mathrm{sc}$.}
\label{FIG_Phase_Cont}
\end{figure*}

\section{Discussion of the Continuum Model}
\label{SFS2D::sec:B}

In this section, we analyze the junction using the continuum model given by Eqs.~(\ref{SFS2D::eq:1}--\ref{mu_continuum_model}). A comparison of the dispersion relations, see Figs.~\ref{FIG_ThreeRegimes} and \ref{FIG_ThreeRegimes_Cont}, shows that the continuum and tight-binding models are in qualitative agreement. A more detailed discussion is provided at the end of Sec.~\ref{SFS2D::sec:3}. The behavior of the energy-phase relation, see Figs.~\ref{FIG_ContinuumModelResults} and \ref{FIG_Phase_Cont}, is also robust to the choice of model. As seen in panel (a), for small spin splitting the junction is in the 0 phase regardless of its transverse momentum $k_y$. By increasing the spin splitting, the junction enters an intermediate regime characterized by qualitatively different behavior as function of $k_y$, see panel (b). Finally, the junction enters the  $\pi$ phase at large values of $\lambda_\mathrm{M}$. 

In comparing predictions of the two models, it is important to choose parameters which are in corresponding physical regimes. This point is not so trivial, and in the rest of this section we discuss how to relate the two sets of parameters. Consider first the ferromagnet-superconductor interface, where in the tight-binding model we have introduced a ferromagnet-superconductor tunneling amplitude $t_c$. Instead, in the continuum model the interface is described by $\delta$-function tunnel barriers of strength $l_0V_0$ [see Eq.~(\ref{Vb})]. To relate $t_c$ and $l_0V_0$, we consider a 1D toy model of a single interface: 
\begin{align}\label{Hcont1D}
H_{\text{cont}} = -\frac{\hbar^2}{2m} \frac{\partial^2}{\partial x^2}- \mu(x) + l_0V_0 \delta(x),
\end{align}
where $l_0V_0 > 0 $ and $\mu(x) = \mu_L \theta(-x) +\mu_R \theta(x)$. As we wish to establish an approximate relation, here we have neglected superconductivity and ferromagnetism. The momentum $k_y$ along the interface is conserved, thus does not appear explicitly (it affects the values of $\mu_{L,R}$). At energy $E$, Eq.~(\ref{Hcont1D}) gives the tunneling coefficient:
\begin{align}
T_{\text{cont}} = \frac{4k_L k_R}{ \left( k_L + k_R \right)^2 + \lambda_0^2},
\end{align}
where $k_{L/R} = \sqrt{2m\left( E + \mu_{L/R} \right)}/\hbar$ and $\lambda_0 = 2ml_0V_0/\hbar^2$. On the other hand, the 1D tight-binding model:
\begin{align}
H_{\text{TB}} = & \sum_{j<1} \epsilon_{L} \hat{c}^{\dagger}_j \hat{c}_j+ \sum_{j\geq 1} \epsilon_{R} \hat{c}^{\dagger}_j \hat{c}_j  -\frac{t_c}{2} \left( \hat{c}^{\dagger}_{1} \hat{c}_{0} + \text{h.c.}\right) \nonumber \\ 
&-\frac{t}{2}\sum_{j\neq 0} \left( \hat{c}^{\dagger}_{j+1} \hat{c}_{j} + \text{h.c.} \right),
\end{align}
gives the tunneling coefficient
\begin{align}
T_{\text{TB}} = \frac{ 4 \sin k_{L}^{\text{(TB)}} \sin k_{R}^{\text{(TB)}}}{ 2 \left[ 1 - \cos\left( k_{L}^{\text{(TB)}} + k_{R}^{\text{(TB)}} \right) \right] + \left( \frac{t_c}{t} - \frac{t}{t_c} \right)^2},
\end{align}
with $k_{L/R}^{\text{(TB)}} = \arccos \left[ -\left( E-\epsilon_{L/R}\right)/t	 \right]$. After taking the long-wavelength limit, i.e., setting $\sin k_{L/R}^{(\text{TB})} \approx k_{L/R}^{(\text{TB})}  = a k_{L/R}$, the expressions of $T_{\text{cont}}$ and $T_{\text{TB}}$ coincide, giving:
\begin{align}
\frac{2ml_0V_0}{\hbar^2} \simeq  \frac{1}{a} \left| \frac{t}{t_c} - \frac{t_c}{t} \right|	
.\end{align}
By dividing both sides by $k_F = \sqrt{ 2m \bar{\mu}_\mathrm{sc} }/\hbar$ and assuming $a k_F  \approx \pi$, we obtain Eq.~\eqref{SFS2D::eq:2} in the main text.

%\begin{align}
%\frac{1}{k_F} \frac{2ml_0V_0}{\hbar^2} = 
%\sqrt{ \frac{2m}{\mu_{sc}} } \frac{l_0V_0}{\hbar} = 
%\frac{1}{k_F a_{\text{T.B.}} }
%\left(\frac{t_{\mathrm{sc}}}{t_\mathrm{c}}
%-\frac{t_\mathrm{c}}{t_{\mathrm{sc}}}\right) 
%\approx \frac{1}{\pi}
%\left(\frac{t_{\mathrm{sc}}}{t_\mathrm{c}}
%-\frac{t_\mathrm{c}}{t_{\mathrm{sc}}}\right) 
%\end{align}
%\begin{align}
%\frac{1}{\pi}
%\left(\frac{t_{\mathrm{sc}}}{t_\mathrm{c}}
%-\frac{t_\mathrm{c}}{t_{\mathrm{sc}}}\right) &
%\approx \sqrt{\frac{2m}{\mu_\mathrm{sc}}}\frac{aV_{0}}{\hbar}
%\end{align}

%
%
Another aspect worth discussing is the choice of $\bar{\mu}_\mathrm{fm}$. This parameter is critical in determining if the ferromagnet is insulating or half-metallic. However (differently from $\mu_\mathrm{fm}$, in the tight-biding model), $\bar{\mu}_\mathrm{fm}$ should take into account the sizable confinement energy induced by the narrow ferromagnetic strip. For an opaque interface, the half-metallic or insulating behavior is determined by the energy $E$ of the lowest 1D subband (at $k_y=0$), rather than directly on $\bar{\mu}_\mathrm{fm}$. The value of $E$ depends on the chemical potential $\bar{\mu}_\mathrm{fm}$, but also on the width $L$ of the strip and the tunnel barriers.

To estimate a suitable $\bar{\mu}_\mathrm{fm}$, we suppose that the lowest subband energy of the isolated ferromagnet is within the superconducting gap at $k_y=0$, and corresponds to the desired behavior (say, insulating). At such energy $E$, the exponential decay of the wavefucntion in the superconductor is given by
\begin{equation}
\kappa = \text{Im}\left[ \sqrt{\frac{2m}{\hbar^2}}  \sqrt{\bar\mu_{\mathrm{sc}} + \sqrt{E^2 - \Delta_{\mathrm{sc}}^2 }}\right].
\end{equation} 
As we are interested in the lowest-energy 1D subband, we consider a nodeless wavefunction of the following form:
\begin{align}
\psi(x) = \left\{ 
\begin{array}{cl}
c_\mathrm{sc} e^{-\kappa|x|} & \mathrm{for}~|x|>L/2, \\
c_{\text{fm}} \cos(kx) & \mathrm{for}~|x|\leq L/2. 
\end{array}
\right.
\end{align}
Here, for simplicity, we have neglected superconductivity and magnetism and assumed a single-component wavefuction. The wavevector inside the strip is given by $k=\sqrt{2m(E-\bar{\mu}_\mathrm{fm})}/\hbar$. Note that the confinement inside the strip leads to an increase in energy, therefore $E$ should satisfy $E-\bar{\mu}_\mathrm{fm}>0$. For given $E$, we compute $c_\mathrm{sc, fm}$ and $\bar{\mu}_\mathrm{fm}$ from the interface potential $V_b(x)$, by imposing the usual boundary conditions of the $\delta$ functions at $x=\pm L/2$.

In summary, using a desired (approximate) subband energy $E$ as an input, the above procedure allows us to find a suitable value of $\bar{\mu}_\mathrm{fm}$, which is the actual parameter entering Eq.~(\ref{BdG_Hamiltonian}). We can consider $\bar\mu_\mathrm{fm}^\mathrm{eff}=-E$ as an effective chemical potential, corresponding more closely to $\mu_\mathrm{fm}$ of the tight-binding model. Fixing $\bar\mu_\mathrm{fm}^\mathrm{eff}$ instead of $\bar{\mu}_\mathrm{fm}$ is especially useful when changing $l_0 V_0$ continuously, as in panels (c) and (f) of Fig.~\ref{FIG_ThreeRegimes_Cont}. This approach allows us to increase the leakage in the superconductor, thus decreasing the confinement energy, without changing the physical regime of interest (insulating or half-metallic).

\section{Proximity Effect on the Ferromagnet}
\label{SFS2D::sec:A}

In this appendix, we derive the effective model for the ferromagnet alone within the subgap regime, starting from the partition function of the system.
\cite{2012_RepProgPhys_Alicea} In the functional-integral representation it is given as follows:
% Since the physical properties at the low energy are related to the
% behaviors of the sub-gap states ($|E|\leq \Delta_{\mathrm{sc}}$),
% in this regard it is often useful to derive an effective description
% for the ferromagnet alone. This could be achieved by integrating out
% the degree of freedoms associated with the SCs. Following the theoretical
% treatments of the proximity effect
% we wrote the the partition function for the full system in term of
% the imaginary-time path integral representation, 
\begin{equation}
\mathcal{Z}=\int\mathcal{D}[c]\mathcal{D}[f]\exp[-\mathcal{S}],\label{partition_function}
\end{equation}
with the total action
\begin{equation}
\calS = \calS_\mathrm{L} + \calS_\mathrm{R} + \calS_\mathrm{fm}
+ \calS_\mathrm{tun},
\end{equation}
where the terms correspond to the respective terms in the total Hamiltonian in Eq.~\eqref{Hamiltonian}.
\begin{widetext}
Specifically, the components of the total action are given by
\begin{subequations}
\begin{align}
\label{total_action}
\calS_\mathrm{L} &=\sum_n\sum_{y,y'} \sum_{x,x'<0}
\mathcal{C}_{x,y,n}^{\dagger}
\left[-\mathcal{G}_{\mathrm{L}}^{-1}(i\omega_{n};x,x',y-y')\right]
\mathcal{C}_{x',y',n} \\
\calS_\mathrm{R} &=\sum_n\sum_{y,y'}  \sum_{x,x'>0}
\mathcal{C}_{x,y,n}^{\dagger}
\left[-\mathcal{G}_{\mathrm{R}}^{-1}(i\omega_{n};x,x',y-y')\right]
\mathcal{C}_{x',y',n} \\
\calS_\mathrm{fm} &= \sum_{n} \sum_{y,y'}
\mathcal{F}_{y,n}^{\dagger}
\left[-\mathcal{G}_{\mathrm{fm}}^{-1}(i\omega_{n};y-y')\right]
\mathcal{F}_{y',n} \\
\calS_\mathrm{tun} &= 
-\frac{t_\mathrm{c}}{2}\sum_n\sum_{y}
\left(\mathcal{C}_{-1,y,n}^{\dagger}\tau^{z}\mathcal{F}_{y,n}
+ \mathcal{C}_{+1,y,n}^{\dagger}\tau^{z}\mathcal{F}_{y,n}
+ \mathrm{h.c.}\right)\color{black}
\end{align}
\end{subequations}
with Green functions
\begin{align}
\calG_\mathrm{L/R}^{-1}(i\omega_n;x,x',y-y') &=
i\omega_n - \calH_\mathrm{L/R}(x,x',y-y') \,, \\
\calG_\mathrm{fm}^{-1}(i\omega_n;x,x',y-y') &=
i\omega_n - \calH_\mathrm{fm}(x,x',y-y') \,,
\end{align}
where
\begin{math}
\omega_n:=2\pi(n+1/2)/\beta
\end{math},
for integer $n$ and inverse temperature $\beta$, is the Fermion Matsubara
frequency and $\calH_\mathrm{L/R}$ and $\calH_\mathrm{fm}$ are single-particle
Hamiltonians for the superconductors and ferromagnet, respectively,
corresponding to Eqs.~\eqref{SC_Hamiltonian} and \eqref{SFS2D::eq:3}.
\end{widetext}
% Here, the first line of Eq.~(\ref{total_action} represents the isolated
% ferromagnet, the second and the third lines for the left and right SCs
% respectively, the last line comes from the tunneling coupling between the
% SCs and the ferromagnet. $\omega_{n}$s are the Matsubara frequencies.
Grassmann-valued fields $\mathcal{C}_{x,y,n}$ and $\mathcal{F}_{y,n}$ take  the same Nambu form as in Eq.~\eqref{Nambu_convention}:
\begin{align}
\label{Nambu_spinor}
\mathcal{C}_{x,y,n} &= \left[\begin{array}{cccc}
c_{x,y,n,\uparrow} & c_{x,y,n,\downarrow} & c_{x,y,\downarrow}^{*} & -c_{x,y,\uparrow}^{*}\end{array}\right]^{\mathrm{T}}, \\
\mathcal{F}_{y,n} &= \left[\begin{array}{cccc}
f_{y,n,\uparrow} & f_{y,n,\downarrow} & f_{y,\downarrow}^{*} & -f_{y,\uparrow}^{*}\end{array}\right]^{\mathrm{T}}.
\end{align}

The effective action for the ferromagnet is obtained by integrating
out the $c$-field from the action in Eq.~\eqref{partition_function} and reads as
\begin{equation}
\mathcal{S}_{\mathrm{eff}}
=\sum_{y,y'}\sum_{n}\mathcal{F}_{y,n}^{\dagger}
[-\mathcal{G}_{\mathrm{eff}}^{-1}(i\omega_{n};y-y')]\mathcal{F}_{y',n}
\end{equation}
with the effective Green's function
\begin{equation}
\mathcal{G}_{\mathrm{eff}}^{-1}(i\omega_{n};y-y')
= i\omega_{n}-\mathcal{H}_{\mathrm{eff}}(y-y'),
\end{equation}
where
%\color{red}
\begin{align}
\label{effective_Hamiltonian}
\mathcal{H}_{\mathrm{eff}}(i\omega_{n};& y-y')
= \mathcal{H}_{\mathrm{fm}}(y-y') \nonumber \\
& + \frac{t_\mathrm{c}^{2}}{4}
\tau^{z}\calG_\mathrm{L}(i\omega_{n};-1,-1,y-y')\tau^{z} \nonumber \\
& + \frac{t_\mathrm{c}^{2}}{4}
\tau^{z}\calG_\mathrm{R}(i\omega_{n};1,1,y-y')\tau^{z} .
\end{align}
In the low-energy limit, such that $|\hbar\omega_{n}|\ll\Delta_{\mathrm{sc}}$,
the $\omega_{n}$ dependence in $\mathcal{H}_{\mathrm{eff}}$ can
be ignored. Furthermore, considering wave-vectors close to the Fermi wave number, such that
\begin{math}
\epsilon_{\mathbf{k}}\approx\mu_{\mathrm{sc}},
\end{math}
Eq.~\eqref{effective_Hamiltonian} in momentum space becomes
%\color{red}
\begin{align}
\label{effective_Hamiltonian_approximated}
\mathcal{H}_{\mathrm{eff}}\approx &
\left[\lambda_{\mathrm{M}}\sigma^{z}-t_\mathrm{fm}\cos(k_y a)
-\mu_{\mathrm{fm}}\right]\tau^{z} \nonumber \\
& + \tau^{x}\mathrm{Re}\Delta_{\mathrm{eff}}
-\tau^{y}\mathrm{Im}\Delta_{\mathrm{eff}},
\end{align}
%\color{black}
with lattice constant $a$ and the proximity-induced pairing potential $\Delta_\mathrm{eff}$ as in Eq.~\eqref{SFS2D::eq:4}. This single-particle form of the effective BdG Hamiltonian for the ferromagnet alone is equivalent to Eq.~\eqref{SFS2D::eq:5} of the main text, where it is expressed in terms of electron annihilation and creation operators, $\hatf_{y,\sigma}$ and $\hatf_{y,\sigma}^\dag$.

\bibliographystyle{apsrev}
\bibliography{References}

\begin{thebibliography}{47}
\expandafter\ifx\csname natexlab\endcsname\relax\def\natexlab#1{#1}\fi
\expandafter\ifx\csname bibnamefont\endcsname\relax
  \def\bibnamefont#1{#1}\fi
\expandafter\ifx\csname bibfnamefont\endcsname\relax
  \def\bibfnamefont#1{#1}\fi
\expandafter\ifx\csname citenamefont\endcsname\relax
  \def\citenamefont#1{#1}\fi
\expandafter\ifx\csname url\endcsname\relax
  \def\url#1{\texttt{#1}}\fi
\expandafter\ifx\csname urlprefix\endcsname\relax\def\urlprefix{URL }\fi
\providecommand{\bibinfo}[2]{#2}
\providecommand{\eprint}[2][]{\url{#2}}

\bibitem[{\citenamefont{Yu}(1965)}]{1965_Acta_Phys_Sin_Yu}
\bibinfo{author}{\bibfnamefont{L.}~\bibnamefont{Yu}}, \bibinfo{journal}{Acta.
  Phys. Sin.} \textbf{\bibinfo{volume}{21}}, \bibinfo{pages}{75}
  (\bibinfo{year}{1965}).

\bibitem[{\citenamefont{Shiba}(1968)}]{1968_Prog_Theor_Phys_Shiba}
\bibinfo{author}{\bibfnamefont{H.}~\bibnamefont{Shiba}},
  \bibinfo{journal}{Prog. Theor. Phys.} \textbf{\bibinfo{volume}{40}},
  \bibinfo{pages}{435} (\bibinfo{year}{1968}).

\bibitem[{\citenamefont{Rusinov}(1969)}]{1969_JETP_Lett_Rusinov}
\bibinfo{author}{\bibfnamefont{A.~I.} \bibnamefont{Rusinov}},
  \bibinfo{journal}{JETP Lett.} \textbf{\bibinfo{volume}{9}},
  \bibinfo{pages}{85} (\bibinfo{year}{1969}).

\bibitem[{\citenamefont{Flude}(1964)}]{1964_PR_Flude}
\bibinfo{author}{\bibfnamefont{R.~A.} \bibnamefont{Flude},
  \bibfnamefont{P.and~Ferrell}}, \bibinfo{journal}{Phys. Rev.}
  \textbf{\bibinfo{volume}{135}}, \bibinfo{pages}{A550} (\bibinfo{year}{1964}).

\bibitem[{\citenamefont{Larkin and Ovchinnikov}(1965)}]{1965_JETP_Larkin}
\bibinfo{author}{\bibfnamefont{A.~I.} \bibnamefont{Larkin}} \bibnamefont{and}
  \bibinfo{author}{\bibfnamefont{Y.~N.} \bibnamefont{Ovchinnikov}},
  \bibinfo{journal}{Sov. Phys. JETP} \textbf{\bibinfo{volume}{20}},
  \bibinfo{pages}{762} (\bibinfo{year}{1965}).

\bibitem[{\citenamefont{He et~al.}(2017)\citenamefont{He, Pan, Stern, Burks,
  Che, Yin, Wang, Lian, Zhou, Choi et~al.}}]{He17a}
\bibinfo{author}{\bibfnamefont{Q.~L.} \bibnamefont{He}},
  \bibinfo{author}{\bibfnamefont{L.}~\bibnamefont{Pan}},
  \bibinfo{author}{\bibfnamefont{A.~L.} \bibnamefont{Stern}},
  \bibinfo{author}{\bibfnamefont{E.~C.} \bibnamefont{Burks}},
  \bibinfo{author}{\bibfnamefont{X.}~\bibnamefont{Che}},
  \bibinfo{author}{\bibfnamefont{G.}~\bibnamefont{Yin}},
  \bibinfo{author}{\bibfnamefont{J.}~\bibnamefont{Wang}},
  \bibinfo{author}{\bibfnamefont{B.}~\bibnamefont{Lian}},
  \bibinfo{author}{\bibfnamefont{Q.}~\bibnamefont{Zhou}},
  \bibinfo{author}{\bibfnamefont{E.~S.} \bibnamefont{Choi}},
  \bibnamefont{et~al.}, \bibinfo{journal}{Science}
  \textbf{\bibinfo{volume}{357}}, \bibinfo{pages}{294} (\bibinfo{year}{2017}).

\bibitem[{\citenamefont{Qi et~al.}(2010)\citenamefont{Qi, Hughes, and
  Zhang}}]{Qi10b}
\bibinfo{author}{\bibfnamefont{X.-L.} \bibnamefont{Qi}},
  \bibinfo{author}{\bibfnamefont{T.~L.} \bibnamefont{Hughes}},
  \bibnamefont{and} \bibinfo{author}{\bibfnamefont{S.-C.} \bibnamefont{Zhang}},
  \bibinfo{journal}{Phys. Rev. B} \textbf{\bibinfo{volume}{82}},
  \bibinfo{pages}{184516} (\bibinfo{year}{2010}).

\bibitem[{\citenamefont{Eschrig}(2011)}]{2011_Phys_Today_Eschrig}
\bibinfo{author}{\bibfnamefont{M.}~\bibnamefont{Eschrig}},
  \bibinfo{journal}{Phys. Today} \textbf{\bibinfo{volume}{64}},
  \bibinfo{pages}{43} (\bibinfo{year}{2011}).

\bibitem[{\citenamefont{Kouwenhoven and Glazman}(2001)}]{Kouwenhoven01a}
\bibinfo{author}{\bibfnamefont{L.}~\bibnamefont{Kouwenhoven}} \bibnamefont{and}
  \bibinfo{author}{\bibfnamefont{L.}~\bibnamefont{Glazman}},
  \bibinfo{journal}{Physics World} \textbf{\bibinfo{volume}{14}},
  \bibinfo{pages}{33} (\bibinfo{year}{2001}).

\bibitem[{\citenamefont{Kondo}(1964)}]{1964_ProgTheorPhys_Kondo}
\bibinfo{author}{\bibfnamefont{J.}~\bibnamefont{Kondo}},
  \bibinfo{journal}{Prog. Theor. Phys.} \textbf{\bibinfo{volume}{32}},
  \bibinfo{pages}{37} (\bibinfo{year}{1964}).

\bibitem[{\citenamefont{Wilson}(1975)}]{1975_RMP_Wilson}
\bibinfo{author}{\bibfnamefont{K.~G.} \bibnamefont{Wilson}},
  \bibinfo{journal}{Rev. Mod. Phys.} \textbf{\bibinfo{volume}{47}},
  \bibinfo{pages}{773} (\bibinfo{year}{1975}).

\bibitem[{\citenamefont{Hewson}(1993)}]{1993_BOOK_Hewson}
\bibinfo{author}{\bibfnamefont{A.~C.} \bibnamefont{Hewson}},
  \emph{\bibinfo{title}{The Kondo Problem to Heavy Fermions}}
  (\bibinfo{publisher}{Cambridge University Press, New York},
  \bibinfo{year}{1993}).

\bibitem[{\citenamefont{Sakurai}(1970)}]{1970_ProgTheorPhys_Sakurai}
\bibinfo{author}{\bibfnamefont{A.}~\bibnamefont{Sakurai}},
  \bibinfo{journal}{Prog. Theor. Phys.} \textbf{\bibinfo{volume}{44}},
  \bibinfo{pages}{1472} (\bibinfo{year}{1970}).

\bibitem[{\citenamefont{Balatsky et~al.}(2006)\citenamefont{Balatsky, Vekhter,
  and Zhu}}]{2006_RMP_Balatsky}
\bibinfo{author}{\bibfnamefont{A.~V.} \bibnamefont{Balatsky}},
  \bibinfo{author}{\bibfnamefont{I.}~\bibnamefont{Vekhter}}, \bibnamefont{and}
  \bibinfo{author}{\bibfnamefont{J.-X.} \bibnamefont{Zhu}},
  \bibinfo{journal}{Rev. Mod. Phys.} \textbf{\bibinfo{volume}{78}},
  \bibinfo{pages}{373} (\bibinfo{year}{2006}),
  \urlprefix\url{https://link.aps.org/doi/10.1103/RevModPhys.78.373}.

\bibitem[{\citenamefont{Choi et~al.}(2004)\citenamefont{Choi, Lee, Kang, and
  Belzig}}]{2004_PRB_Choi}
\bibinfo{author}{\bibfnamefont{M.-S.} \bibnamefont{Choi}},
  \bibinfo{author}{\bibfnamefont{M.}~\bibnamefont{Lee}},
  \bibinfo{author}{\bibfnamefont{K.}~\bibnamefont{Kang}}, \bibnamefont{and}
  \bibinfo{author}{\bibfnamefont{W.}~\bibnamefont{Belzig}},
  \bibinfo{journal}{Phys. Rev. B} \textbf{\bibinfo{volume}{70}},
  \bibinfo{pages}{020502} (\bibinfo{year}{2004}),
  \urlprefix\url{https://link.aps.org/doi/10.1103/PhysRevB.70.020502}.

\bibitem[{\citenamefont{\ifmmode~\check{Z}\else \v{Z}\fi{}itko
  et~al.}(2010)\citenamefont{\ifmmode~\check{Z}\else \v{Z}\fi{}itko, Lee,
  L\'opez, Aguado, and Choi}}]{2010_PRL_Zitko}
\bibinfo{author}{\bibfnamefont{R.}~\bibnamefont{\ifmmode~\check{Z}\else
  \v{Z}\fi{}itko}}, \bibinfo{author}{\bibfnamefont{M.}~\bibnamefont{Lee}},
  \bibinfo{author}{\bibfnamefont{R.}~\bibnamefont{L\'opez}},
  \bibinfo{author}{\bibfnamefont{R.}~\bibnamefont{Aguado}}, \bibnamefont{and}
  \bibinfo{author}{\bibfnamefont{M.-S.} \bibnamefont{Choi}},
  \bibinfo{journal}{Phys. Rev. Lett.} \textbf{\bibinfo{volume}{105}},
  \bibinfo{pages}{116803} (\bibinfo{year}{2010}),
  \urlprefix\url{https://link.aps.org/doi/10.1103/PhysRevLett.105.116803}.

\bibitem[{\citenamefont{Andreev et~al.}(1991)\citenamefont{Andreev, Buzdin, and
  Osgood}}]{1991_PRB_Andreev}
\bibinfo{author}{\bibfnamefont{A.~V.} \bibnamefont{Andreev}},
  \bibinfo{author}{\bibfnamefont{A.~I.} \bibnamefont{Buzdin}},
  \bibnamefont{and} \bibinfo{author}{\bibfnamefont{R.~M.}
  \bibnamefont{Osgood}}, \bibinfo{journal}{Phys. Rev. B}
  \textbf{\bibinfo{volume}{43}}, \bibinfo{pages}{10124} (\bibinfo{year}{1991}),
  \urlprefix\url{https://link.aps.org/doi/10.1103/PhysRevB.43.10124}.

\bibitem[{\citenamefont{Bulaevskii et~al.}(1977)\citenamefont{Bulaevskii,
  Kuzii, and Sobyanin}}]{1977_JETPLett_Bulaevskii}
\bibinfo{author}{\bibfnamefont{L.~N.} \bibnamefont{Bulaevskii}},
  \bibinfo{author}{\bibfnamefont{V.~V.} \bibnamefont{Kuzii}}, \bibnamefont{and}
  \bibinfo{author}{\bibfnamefont{A.~A.} \bibnamefont{Sobyanin}},
  \bibinfo{journal}{JETP Lett.} \textbf{\bibinfo{volume}{25}},
  \bibinfo{pages}{314} (\bibinfo{year}{1977}).

\bibitem[{\citenamefont{Buzdin et~al.}(1982)\citenamefont{Buzdin, Bulaevskii,
  and Panyukov}}]{1982_JETPLett_Buzdin}
\bibinfo{author}{\bibfnamefont{A.~I.} \bibnamefont{Buzdin}},
  \bibinfo{author}{\bibfnamefont{L.~N.} \bibnamefont{Bulaevskii}},
  \bibnamefont{and} \bibinfo{author}{\bibfnamefont{S.~V.}
  \bibnamefont{Panyukov}}, \bibinfo{journal}{JETP Lett}
  \textbf{\bibinfo{volume}{35}}, \bibinfo{pages}{147} (\bibinfo{year}{1982}).

\bibitem[{\citenamefont{Ryazanov et~al.}(2001)\citenamefont{Ryazanov, Oboznov,
  Rusanov, Veretennikov, Golubov, and Aarts}}]{2001_PRL_Ryazanov_Coupling}
\bibinfo{author}{\bibfnamefont{V.~V.} \bibnamefont{Ryazanov}},
  \bibinfo{author}{\bibfnamefont{V.~A.} \bibnamefont{Oboznov}},
  \bibinfo{author}{\bibfnamefont{A.~Y.} \bibnamefont{Rusanov}},
  \bibinfo{author}{\bibfnamefont{A.~V.} \bibnamefont{Veretennikov}},
  \bibinfo{author}{\bibfnamefont{A.~A.} \bibnamefont{Golubov}},
  \bibnamefont{and} \bibinfo{author}{\bibfnamefont{J.}~\bibnamefont{Aarts}},
  \bibinfo{journal}{Phys. Rev. Lett.} \textbf{\bibinfo{volume}{86}},
  \bibinfo{pages}{2427} (\bibinfo{year}{2001}).

\bibitem[{\citenamefont{Costa et~al.}(2018)\citenamefont{Costa, Fabian, and
  Kochan}}]{2018_PRB_Costa}
\bibinfo{author}{\bibfnamefont{A.}~\bibnamefont{Costa}},
  \bibinfo{author}{\bibfnamefont{J.}~\bibnamefont{Fabian}}, \bibnamefont{and}
  \bibinfo{author}{\bibfnamefont{D.}~\bibnamefont{Kochan}},
  \bibinfo{journal}{Phys. Rev. B} \textbf{\bibinfo{volume}{98}},
  \bibinfo{pages}{134511} (\bibinfo{year}{2018}),
  \urlprefix\url{https://link.aps.org/doi/10.1103/PhysRevB.98.134511}.

\bibitem[{\citenamefont{Nadj-Perge et~al.}(2014)\citenamefont{Nadj-Perge,
  Drozdov, Li, Chen, Jeon, Seo, MacDonald, Bernevig, and
  Yazdani}}]{2014_Science_NadjPerge}
\bibinfo{author}{\bibfnamefont{S.}~\bibnamefont{Nadj-Perge}},
  \bibinfo{author}{\bibfnamefont{I.~K.} \bibnamefont{Drozdov}},
  \bibinfo{author}{\bibfnamefont{J.}~\bibnamefont{Li}},
  \bibinfo{author}{\bibfnamefont{H.}~\bibnamefont{Chen}},
  \bibinfo{author}{\bibfnamefont{S.}~\bibnamefont{Jeon}},
  \bibinfo{author}{\bibfnamefont{J.}~\bibnamefont{Seo}},
  \bibinfo{author}{\bibfnamefont{A.~H.} \bibnamefont{MacDonald}},
  \bibinfo{author}{\bibfnamefont{B.~A.} \bibnamefont{Bernevig}},
  \bibnamefont{and} \bibinfo{author}{\bibfnamefont{A.}~\bibnamefont{Yazdani}},
  \bibinfo{journal}{Science} \textbf{\bibinfo{volume}{346}},
  \bibinfo{pages}{602} (\bibinfo{year}{2014}).

\bibitem[{\citenamefont{Peng et~al.}(2015)\citenamefont{Peng, Pientka, Glazman,
  and von Oppen}}]{2015_PRL_Peng}
\bibinfo{author}{\bibfnamefont{Y.}~\bibnamefont{Peng}},
  \bibinfo{author}{\bibfnamefont{F.}~\bibnamefont{Pientka}},
  \bibinfo{author}{\bibfnamefont{L.~I.} \bibnamefont{Glazman}},
  \bibnamefont{and} \bibinfo{author}{\bibfnamefont{F.}~\bibnamefont{von
  Oppen}}, \bibinfo{journal}{Phys. Rev. Lett.} \textbf{\bibinfo{volume}{114}},
  \bibinfo{pages}{106801} (\bibinfo{year}{2015}),
  \urlprefix\url{https://link.aps.org/doi/10.1103/PhysRevLett.114.106801}.

\bibitem[{\citenamefont{Hatter et~al.}(2015)\citenamefont{Hatter, Heinrich,
  Ruby, Pascual, and Franke}}]{2015_NatCommun_Hatter}
\bibinfo{author}{\bibfnamefont{N.}~\bibnamefont{Hatter}},
  \bibinfo{author}{\bibfnamefont{B.~W.} \bibnamefont{Heinrich}},
  \bibinfo{author}{\bibfnamefont{M.}~\bibnamefont{Ruby}},
  \bibinfo{author}{\bibfnamefont{J.~I.} \bibnamefont{Pascual}},
  \bibnamefont{and} \bibinfo{author}{\bibfnamefont{K.~J.}
  \bibnamefont{Franke}}, \bibinfo{journal}{Nat. Commun.}
  \textbf{\bibinfo{volume}{6}}, \bibinfo{pages}{8988} (\bibinfo{year}{2015}).

\bibitem[{\citenamefont{Ding et~al.}(2021)\citenamefont{Ding, Hu, Randeria,
  Hoffman, Deb, Klinovaja, Loss, and Yazdani}}]{2021_PNAS_Ding}
\bibinfo{author}{\bibfnamefont{H.}~\bibnamefont{Ding}},
  \bibinfo{author}{\bibfnamefont{Y.}~\bibnamefont{Hu}},
  \bibinfo{author}{\bibfnamefont{M.~T.} \bibnamefont{Randeria}},
  \bibinfo{author}{\bibfnamefont{S.}~\bibnamefont{Hoffman}},
  \bibinfo{author}{\bibfnamefont{O.}~\bibnamefont{Deb}},
  \bibinfo{author}{\bibfnamefont{J.}~\bibnamefont{Klinovaja}},
  \bibinfo{author}{\bibfnamefont{D.}~\bibnamefont{Loss}}, \bibnamefont{and}
  \bibinfo{author}{\bibfnamefont{A.}~\bibnamefont{Yazdani}},
  \bibinfo{journal}{Proc. Natl. Acad. Sci.} \textbf{\bibinfo{volume}{118}},
  \bibinfo{pages}{e2024837118} (\bibinfo{year}{2021}),
  \urlprefix\url{https://www.pnas.org/doi/abs/10.1073/pnas.2024837118}.

\bibitem[{\citenamefont{Onishi et~al.}(1996)\citenamefont{Onishi, Ohashi,
  Shingaki, and Miyake}}]{1996_JPSJ_Onishi}
\bibinfo{author}{\bibfnamefont{Y.}~\bibnamefont{Onishi}},
  \bibinfo{author}{\bibfnamefont{Y.}~\bibnamefont{Ohashi}},
  \bibinfo{author}{\bibfnamefont{Y.}~\bibnamefont{Shingaki}}, \bibnamefont{and}
  \bibinfo{author}{\bibfnamefont{K.}~\bibnamefont{Miyake}},
  \bibinfo{journal}{J. Phys. Soc. Jpn.} \textbf{\bibinfo{volume}{65}},
  \bibinfo{pages}{675} (\bibinfo{year}{1996}).

\bibitem[{\citenamefont{Zhu and Ting}(2000)}]{2000_PRB_Zhu}
\bibinfo{author}{\bibfnamefont{J.-X.} \bibnamefont{Zhu}} \bibnamefont{and}
  \bibinfo{author}{\bibfnamefont{C.~S.} \bibnamefont{Ting}},
  \bibinfo{journal}{Phys. Rev. B} \textbf{\bibinfo{volume}{61}},
  \bibinfo{pages}{1456} (\bibinfo{year}{2000}).

\bibitem[{\citenamefont{Blonder et~al.}(1982)\citenamefont{Blonder, Tinkham,
  and Klapwijk}}]{1982_PRB_Blonder}
\bibinfo{author}{\bibfnamefont{G.~E.} \bibnamefont{Blonder}},
  \bibinfo{author}{\bibfnamefont{M.}~\bibnamefont{Tinkham}}, \bibnamefont{and}
  \bibinfo{author}{\bibfnamefont{T.~M.} \bibnamefont{Klapwijk}},
  \bibinfo{journal}{Phys. Rev. B} \textbf{\bibinfo{volume}{25}},
  \bibinfo{pages}{4515} (\bibinfo{year}{1982}),
  \urlprefix\url{https://link.aps.org/doi/10.1103/PhysRevB.25.4515}.

\bibitem[{\citenamefont{Beenakker}(1991)}]{1991_PRL_Beenakker}
\bibinfo{author}{\bibfnamefont{C.~W.} \bibnamefont{Beenakker}},
  \bibinfo{journal}{Phys. Rev. Lett.} \textbf{\bibinfo{volume}{67}},
  \bibinfo{pages}{3836} (\bibinfo{year}{1991}).

\bibitem[{\citenamefont{de~Gennes}(1966)}]{1966_BOOK_deGennes}
\bibinfo{author}{\bibfnamefont{P.~G.} \bibnamefont{de~Gennes}},
  \emph{\bibinfo{title}{Superconductivity of Metals and Alloys}}
  (\bibinfo{publisher}{Benjamin, New York}, \bibinfo{year}{1966}).

\bibitem[{\citenamefont{Zyuzin and Loss}(2014)}]{2014_PRB_Zyuzin}
\bibinfo{author}{\bibfnamefont{A.~A.} \bibnamefont{Zyuzin}} \bibnamefont{and}
  \bibinfo{author}{\bibfnamefont{D.}~\bibnamefont{Loss}},
  \bibinfo{journal}{Phys. Rev. B} \textbf{\bibinfo{volume}{90}},
  \bibinfo{pages}{125443} (\bibinfo{year}{2014}),
  \urlprefix\url{https://link.aps.org/doi/10.1103/PhysRevB.90.125443}.

\bibitem[{\citenamefont{Deb et~al.}(2021)\citenamefont{Deb, Hoffman, Loss, and
  Klinovaja}}]{2021_PRB_Deb}
\bibinfo{author}{\bibfnamefont{O.}~\bibnamefont{Deb}},
  \bibinfo{author}{\bibfnamefont{S.}~\bibnamefont{Hoffman}},
  \bibinfo{author}{\bibfnamefont{D.}~\bibnamefont{Loss}}, \bibnamefont{and}
  \bibinfo{author}{\bibfnamefont{J.}~\bibnamefont{Klinovaja}},
  \bibinfo{journal}{Phys. Rev. B} \textbf{\bibinfo{volume}{103}},
  \bibinfo{pages}{165403} (\bibinfo{year}{2021}),
  \urlprefix\url{https://link.aps.org/doi/10.1103/PhysRevB.103.165403}.

\bibitem[{\citenamefont{Rouco et~al.}(2019)\citenamefont{Rouco, Tokatly, and
  Bergeret}}]{2019_PRB_Rouco}
\bibinfo{author}{\bibfnamefont{M.}~\bibnamefont{Rouco}},
  \bibinfo{author}{\bibfnamefont{I.~V.} \bibnamefont{Tokatly}},
  \bibnamefont{and} \bibinfo{author}{\bibfnamefont{F.~S.}
  \bibnamefont{Bergeret}}, \bibinfo{journal}{Phys. Rev. B}
  \textbf{\bibinfo{volume}{99}}, \bibinfo{pages}{094514}
  (\bibinfo{year}{2019}).

\bibitem[{\citenamefont{Buzdin}(2005)}]{Buzdin05a}
\bibinfo{author}{\bibfnamefont{A.~I.} \bibnamefont{Buzdin}},
  \bibinfo{journal}{Rev. Mod. Phys.} \textbf{\bibinfo{volume}{77}},
  \bibinfo{pages}{935} (\bibinfo{year}{2005}).

\bibitem[{\citenamefont{Visani et~al.}(2012)\citenamefont{Visani, Sefrioui,
  Tornos, Leon, Briatico, Bibes, Barth{\'e}l{\'e}my, Santamar{\'\i}a, and
  Villegas}}]{Visani12a}
\bibinfo{author}{\bibfnamefont{C.}~\bibnamefont{Visani}},
  \bibinfo{author}{\bibfnamefont{Z.}~\bibnamefont{Sefrioui}},
  \bibinfo{author}{\bibfnamefont{J.}~\bibnamefont{Tornos}},
  \bibinfo{author}{\bibfnamefont{C.}~\bibnamefont{Leon}},
  \bibinfo{author}{\bibfnamefont{J.}~\bibnamefont{Briatico}},
  \bibinfo{author}{\bibfnamefont{M.}~\bibnamefont{Bibes}},
  \bibinfo{author}{\bibfnamefont{A.}~\bibnamefont{Barth{\'e}l{\'e}my}},
  \bibinfo{author}{\bibfnamefont{J.}~\bibnamefont{Santamar{\'\i}a}},
  \bibnamefont{and} \bibinfo{author}{\bibfnamefont{J.~E.}
  \bibnamefont{Villegas}}, \bibinfo{journal}{Nature Physics}
  \textbf{\bibinfo{volume}{8}}, \bibinfo{pages}{539} (\bibinfo{year}{2012}).

\bibitem[{\citenamefont{Eschrig and L{\"o}fwander}(2008)}]{Eschrig08a}
\bibinfo{author}{\bibfnamefont{M.}~\bibnamefont{Eschrig}} \bibnamefont{and}
  \bibinfo{author}{\bibfnamefont{T.}~\bibnamefont{L{\"o}fwander}},
  \bibinfo{journal}{Nature Physics} \textbf{\bibinfo{volume}{4}},
  \bibinfo{pages}{138} (\bibinfo{year}{2008}).

\bibitem[{\citenamefont{Keizer et~al.}(2006)\citenamefont{Keizer, Goennenwein,
  Klapwijk, Miao, Xiao, and Gupta}}]{Keizer06a}
\bibinfo{author}{\bibfnamefont{R.~S.} \bibnamefont{Keizer}},
  \bibinfo{author}{\bibfnamefont{S.~T.~B.} \bibnamefont{Goennenwein}},
  \bibinfo{author}{\bibfnamefont{T.~M.} \bibnamefont{Klapwijk}},
  \bibinfo{author}{\bibfnamefont{G.}~\bibnamefont{Miao}},
  \bibinfo{author}{\bibfnamefont{G.}~\bibnamefont{Xiao}}, \bibnamefont{and}
  \bibinfo{author}{\bibfnamefont{A.}~\bibnamefont{Gupta}},
  \bibinfo{journal}{Nature} \textbf{\bibinfo{volume}{439}},
  \bibinfo{pages}{825} (\bibinfo{year}{2006}).

\bibitem[{\citenamefont{Bratkovsky}(1997)}]{Bratkovsky97a}
\bibinfo{author}{\bibfnamefont{A.~M.} \bibnamefont{Bratkovsky}},
  \bibinfo{journal}{Phys. Rev. B} \textbf{\bibinfo{volume}{56}},
  \bibinfo{pages}{2344} (\bibinfo{year}{1997}).

\bibitem[{\citenamefont{Capogna and Blamire}(1996)}]{Capogna96a}
\bibinfo{author}{\bibfnamefont{L.}~\bibnamefont{Capogna}} \bibnamefont{and}
  \bibinfo{author}{\bibfnamefont{M.~G.} \bibnamefont{Blamire}},
  \bibinfo{journal}{Phys. Rev. B} \textbf{\bibinfo{volume}{53}},
  \bibinfo{pages}{5683} (\bibinfo{year}{1996}).

\bibitem[{\citenamefont{Volkov et~al.}(1995)\citenamefont{Volkov, Magn{\'e}e,
  van Wees, and Klapwijk}}]{Volkov95a}
\bibinfo{author}{\bibfnamefont{A.~F.} \bibnamefont{Volkov}},
  \bibinfo{author}{\bibfnamefont{P.~H.~C.} \bibnamefont{Magn{\'e}e}},
  \bibinfo{author}{\bibfnamefont{B.~J.} \bibnamefont{van Wees}},
  \bibnamefont{and} \bibinfo{author}{\bibfnamefont{T.~M.}
  \bibnamefont{Klapwijk}}, \bibinfo{journal}{Physica C}
  \textbf{\bibinfo{volume}{242}}, \bibinfo{pages}{261} (\bibinfo{year}{1995}).

\bibitem[{\citenamefont{McMillan}(1968)}]{McMillan68a}
\bibinfo{author}{\bibfnamefont{W.~L.} \bibnamefont{McMillan}},
  \bibinfo{journal}{Phys. Rev.} \textbf{\bibinfo{volume}{175}},
  \bibinfo{pages}{537} (\bibinfo{year}{1968}).

\bibitem[{\citenamefont{Mart{\'\i}n-Rodero and
  Levy~Yeyati}(2011)}]{Martin-Rodero11a}
\bibinfo{author}{\bibfnamefont{A.}~\bibnamefont{Mart{\'\i}n-Rodero}}
  \bibnamefont{and}
  \bibinfo{author}{\bibfnamefont{A.}~\bibnamefont{Levy~Yeyati}},
  \bibinfo{journal}{Advances in Physics} \textbf{\bibinfo{volume}{60}},
  \bibinfo{pages}{899} (\bibinfo{year}{2011}).

\bibitem[{\citenamefont{Furusaki et~al.}(1991)\citenamefont{Furusaki,
  Takayanagi, and Tsukada}}]{Furusaki91a}
\bibinfo{author}{\bibfnamefont{A.}~\bibnamefont{Furusaki}},
  \bibinfo{author}{\bibfnamefont{H.}~\bibnamefont{Takayanagi}},
  \bibnamefont{and} \bibinfo{author}{\bibfnamefont{M.}~\bibnamefont{Tsukada}},
  \bibinfo{journal}{Phys. Rev. Lett.} \textbf{\bibinfo{volume}{67}},
  \bibinfo{pages}{132} (\bibinfo{year}{1991}).

\bibitem[{\citenamefont{Caretta et~al.}(2020)\citenamefont{Caretta, Oh,
  Fakhrul, Lee, Lee, Kim, Ross, Lee, and Beach}}]{Caretta20a}
\bibinfo{author}{\bibfnamefont{L.}~\bibnamefont{Caretta}},
  \bibinfo{author}{\bibfnamefont{S.-H.} \bibnamefont{Oh}},
  \bibinfo{author}{\bibfnamefont{T.}~\bibnamefont{Fakhrul}},
  \bibinfo{author}{\bibfnamefont{D.-K.} \bibnamefont{Lee}},
  \bibinfo{author}{\bibfnamefont{B.~H.} \bibnamefont{Lee}},
  \bibinfo{author}{\bibfnamefont{S.~K.} \bibnamefont{Kim}},
  \bibinfo{author}{\bibfnamefont{C.~A.} \bibnamefont{Ross}},
  \bibinfo{author}{\bibfnamefont{K.-J.} \bibnamefont{Lee}}, \bibnamefont{and}
  \bibinfo{author}{\bibfnamefont{G.~S.~D.} \bibnamefont{Beach}},
  \bibinfo{journal}{Science} \textbf{\bibinfo{volume}{370}},
  \bibinfo{pages}{1438} (\bibinfo{year}{2020}).

\bibitem[{\citenamefont{\ifmmode \check{Z}\else
  \v{Z}\fi{}uti\ifmmode~\acute{c}\else \'{c}\fi{}
  et~al.}(2004)\citenamefont{\ifmmode \check{Z}\else
  \v{Z}\fi{}uti\ifmmode~\acute{c}\else \'{c}\fi{}, Fabian, and
  Das~Sarma}}]{Zutick04a}
\bibinfo{author}{\bibfnamefont{I.}~\bibnamefont{\ifmmode \check{Z}\else
  \v{Z}\fi{}uti\ifmmode~\acute{c}\else \'{c}\fi{}}},
  \bibinfo{author}{\bibfnamefont{J.}~\bibnamefont{Fabian}}, \bibnamefont{and}
  \bibinfo{author}{\bibfnamefont{S.}~\bibnamefont{Das~Sarma}},
  \bibinfo{journal}{Rev. Mod. Phys.} \textbf{\bibinfo{volume}{76}},
  \bibinfo{pages}{323} (\bibinfo{year}{2004}).

\bibitem[{\citenamefont{Wolf et~al.}(2001)\citenamefont{Wolf, Awschalom,
  Buhrman, Daughton, von Moln{\'a}r, Roukes, Chtchelkanova, and
  Treger}}]{Wolf01a}
\bibinfo{author}{\bibfnamefont{S.~A.} \bibnamefont{Wolf}},
  \bibinfo{author}{\bibfnamefont{D.~D.} \bibnamefont{Awschalom}},
  \bibinfo{author}{\bibfnamefont{R.~A.} \bibnamefont{Buhrman}},
  \bibinfo{author}{\bibfnamefont{J.~M.} \bibnamefont{Daughton}},
  \bibinfo{author}{\bibfnamefont{S.}~\bibnamefont{von Moln{\'a}r}},
  \bibinfo{author}{\bibfnamefont{M.~L.} \bibnamefont{Roukes}},
  \bibinfo{author}{\bibfnamefont{A.~Y.} \bibnamefont{Chtchelkanova}},
  \bibnamefont{and} \bibinfo{author}{\bibfnamefont{D.~M.}
  \bibnamefont{Treger}}, \bibinfo{journal}{Science}
  \textbf{\bibinfo{volume}{294}}, \bibinfo{pages}{1488} (\bibinfo{year}{2001}).

\bibitem[{\citenamefont{Alicea}(2012)}]{2012_RepProgPhys_Alicea}
\bibinfo{author}{\bibfnamefont{J.}~\bibnamefont{Alicea}},
  \bibinfo{journal}{Rep. Prog. Phys.} \textbf{\bibinfo{volume}{75}},
  \bibinfo{pages}{076501} (\bibinfo{year}{2012}).

\end{thebibliography}

\end{document}